\documentclass[prb,superscriptaddress,showpacs,twocolumn]{revtex4}
\usepackage{graphicx,amsfonts}
\usepackage{epsfig,amsmath,eufrak}

\usepackage{natbib,layout}

\begin{document}

\newcommand{\atanh}
{\operatorname{atanh}}
\newcommand{\ArcTan}
{\operatorname{ArcTan}}
\newcommand{\ArcCoth}
{\operatorname{ArcCoth}}
\newcommand{\Erf}
{\operatorname{Erf}}
\newcommand{\Erfi}
{\operatorname{Erfi}}
\newcommand{\Ei}
{\operatorname{Ei}}

\title{Low Temperature Specific Heat of some Quantum Mean Field glassy
phases.} 
\author{Gregory Schehr}
\affiliation{Theoretische Physik Universit\"at des Saarlandes
66041 Saarbr\"ucken Germany}

\draft

\date{\today}
\begin{abstract}
We investigate analytically the low temperature behavior of the
specific heat $C_v(T)$ for a large class of quantum disordered models 
within Mean Field approximation. This includes the vibrational
modes of a lattice pinned by impurity disorder in the quantum regime,
the quantum spherical $p$-spin-glass and a quantum Heisenberg spin
glass. We exhibit a general mechanism, common to all these models,
arising from the so-called marginality condition, responsible for
the cancellation of the linear and quadratic contributions in $T$ in the
specific heat. We thus find for all these models the Mean Field result
$C_v(T) \propto T^3$. 
\end{abstract}
\maketitle


\newpage

\section{INTRODUCTION.}

While they have been experimentally observed several decades
ago\cite{zeller_chalspe_struct_glasses}, 
the anomalous low temperature thermodynamical
properties of disordered and glassy systems remain a 
formidable theoretical issue. In particular, measurements
of the specific heat $C_v(T)$
in a variety of glasses including structural glasses 
\cite{zeller_chalspe_struct_glasses}, disordered crystals
\cite{ackerman_chalspe_disordered_crystals}, or spin-glasses
\cite{binder_spinglass_review} show a linear behavior $C_v(T)
\propto T$ at low temperature. Such a behavior is
often explained by the standard 
two-level systems (TLS) phenomenological argument
\cite{anderson_twolevels}. Although this TLS argument is very appealing,
and successfull in many situations, it appears extremely hard to
vindicate it from a microscopic description. 

The computation of the specific heat of a disordered system
starting from a microscopic hamiltonian is a
very complicated task. In this respect, important
progress has been achieved by  
recent developments in Mean Field methods in quantum spin-glasses
\cite{ bray_quantum_sg,sachdev_ye_sg,
  georges_mf_quantum_spinglass,cugliandolo_quantum_p_spin} and 
related models \cite{giamarchi_columnar_variat},
allowing for the description of low lying excitations in quantum
glasses. However, even in this solvable limit, the analytical
computation of the $C_v(T)$ is still intricated and the question  
whether this TLS argument is confirmed or not by Mean Field
calculations is still a subject of controversy
\cite{camjayi_sun_sg_num,schehr_chalspe_quantique}.
    
In Ref. \cite{cugliandolo_quantum_p_spin}, a quantum extension of the
spherical $p$-spin model was studied. In the marginal spin-glass phase,
characterized by a one step Replica
Symmetry Breaking (RSB) ansatz together with the marginality
condition, some indications were found for a
linear behavior of the specific heat, 
although its low $T$ behavior was not extracted analytically.
The authors
of Ref.~\cite{georges_mf_quantum_spinglass} have studied a Mean Field theory
of a $SU(2)$ quantum Heisenberg spin-glass. Using a semi-classical
expansion in $1/S$, with $S$ the size of the spins, the specific
heat was obtained analytically to lowest order in the marginal spin
glass phase, also described by a one step RSB solution. At this order,
the linear and quadratic terms of the low 
$T$ expansion of $C_v(T)$ were found to {\it cancel}, leading to a cubic
behavior, $C_v(T) \propto T^3$ (these cancellations were found to
occur in the related model of a quantum Ising spin-glass at the lowest
order in a similar semi-classical expansion
\cite{georges_mf_quantum_spinglass}). The expansion to next order 
appeared to be rather intricated, and it was argued
\cite{georges_mf_quantum_spinglass} that the accidental cancellation
identified to lowest order does not occur to this next order,  
yielding a linear contribution to $C_v(T)$. A later numerical solution
of the saddle point 
equation \cite{camjayi_sun_sg_num} claims instead the absence of this linear
contribution and a low $T$ behavior $C_v(T) \propto T^2$.

A class of models for which such Mean Field methods
have been applied 
with some success, {\it e.g.} to compute correlation functions
\cite{giamarchi_columnar_variat,giamarchi_vortex_long}, are disordered
elastic systems, which cover a wide range of physical situations  
such as charge density waves \cite{gruner_book_cdw},
electron glasses \cite{giamarchi_wigner_review,chitra_wigner_long},   
and flux lattices \cite{blatter_vortex_review,giamarchi_book_young,
  nattermann_vortex_review,giamarchi_vortex_review}, for which the
quantum regime is relevant. In the elastic limit, where topological
defects can be neglected, which is for instance the case in the
so-called Bragg glass phase, these systems have been studied, both in
the classical \cite{giamarchi_vortex_long} and quantum 
\cite{giamarchi_columnar_variat} limit, using
the Gaussian 
variational approximation
\cite{mezard_variational_replica,
  giamarchi_vortex_long,giamarchi_columnar_variat} to the replicated
Hamiltonian. In this framework, the specific heat has been studied
both in the classical \cite{schehr_chalspe_classique} and quantum
regime\cite{schehr_chalspe_quantique,schehr_chalspe_long}. In the
quantum limit, of  
interest here, the specific heat has been computed in a semi
classical expansion in powers of $\hbar$, keeping $\hbar/T$ fixed,
similar to the aforementioned $1/S$ expansion 
\cite{georges_mf_quantum_spinglass}. At the leading order, the
cancellation of the linear and quadratic terms in $C_v(T)$ was also
obtained \cite{schehr_chalspe_quantique,schehr_chalspe_long}. But
surprisingly, the analysis of the next to leading order showed
that these cancellations also occur \cite{schehr_chalspe_long}. 
In view of these results
\cite{georges_mf_quantum_spinglass,camjayi_sun_sg_num,
schehr_chalspe_long}, it is important to know whether there is a
general property, within this Mean Field approach, leading to the
cancellation of the linear term in 
$C_v(T)$.     

In this paper, we identify a general
mechanism, common to all these models, 
relying on the marginality condition, which leads to the cancellation
of the linear and quadratic (in $T$) contribution to $C_v(T)$ at low
$T$. This leads, {\it independently of any semi-classical expansion nor
numerics}, to 
$C_v(T) \propto T^3$. We believe that
this sheds light on the Mean Field approximation applied to these
quantum disordered models. 

The organization of the paper is as follow. In section II, we introduce
the different models we will be interested in, and recall the main
properties of the saddle point equations. Section III is devoted to
the low $T$ expansion itself: we first exhibit the non trivial low $T$
structure of the variational equations, therefore extending the
previous analysis of
Ref.
\cite{georges_mf_quantum_spinglass,cugliandolo_quantum_p_spin,giamarchi_columnar_variat}
at finite $T$, and then turn to the
computation of the specific heat. Finally, we draw our conclusions in
the last section.


\section{MODELS AND MEAN FIELD APPROXIMATIONS.}

\subsection{Quantum periodic elastic manifold in a random potential
  (Model I).} 


We consider a collection of interacting 
quantum objects of mass $M$ whose position variables 
are denoted $u_\alpha(R_i,\tau)$. The 
equilibrium positions $R_i$ in the absence of
any fluctuations form a perfect lattice of spacing $a$.
Interactions result in an elastic tensor $\Phi_{\alpha,\beta}(q)$
which describes the energy associated to small displacements, the
phonon degrees of 
freedom. Impurity disorder is modeled 
by a $\tau$ independent gaussian random potential $U(x)$ 
directly coupled to
the local density $\rho(x) = \sum_i \delta (x - R_i - u(R_i,\tau))$.
We will describe systems in the weak disorder regime $a/R_a \ll 1$
where $R_a$ is the translational correlation length, {\it e.g.} 
in a Bragg glass phase where the condition 
$|u_\alpha(R_i,\tau) - u_\alpha(R_i+a ,\tau)| \ll a$ holds, no
dislocation being present.
The system at equilibrium at temperature $T = 1/\beta$ is described by the
partition function $Z= Tr e^{- \beta H} = \int Du D\Pi e^{-S/\hbar}$
with the Hamiltonian $H=H_{\text{ph}} + H_{\text{dis}}$:
\begin{eqnarray}
&& H_{\text{ph}} = \frac{1}{2} \int_q \frac{\Pi(q)^2}{M} +
  \sum_{\alpha,\beta} u_{\alpha}(q) 
\Phi_{\alpha,\beta}(q) u_{\beta}(-q)  \nonumber \\
&& H_{\text{dis}} =  \int d^dx U(x) \rho(x, u(x)) \label{Hsys}
\end{eqnarray}
and its associated Euclidean quantum action in imaginary time $\tau$
\begin{eqnarray}
-S[\Pi,u]=\int_0^{\beta \hbar} d\tau \int_ q 
i \Pi_{\alpha}(q,\tau) \partial_\tau u_{\alpha}(q,\tau) - H
\label{Def_Action} 
\end{eqnarray}
where
$u(q,\tau)$ and its conjugated momentum $\Pi(q,\tau)$ satisfy periodic
boundary conditions, of period $\beta \hbar$, along the $\tau$
axis. One denotes by  
$\int_q \equiv \int_{BZ} \frac{d^d q}{(2 \pi)^d}$ integration
on the first Brillouin zone. We will focus here on the case of
internal dimension $d \geq 2$. For simplicity we illustrate the
calculation on a 
isotropic system with $\Phi_{\alpha,\beta}(q)= c q^2 \delta_{\alpha 
\beta}$ and denote disorder correlations 
\begin{eqnarray}\label{Dis_Av}
&&\overline{U(x)} = 0 \quad, \quad \overline{U(x) U(x')} = \Delta(x-x')
\end{eqnarray}
The disorder average is performed using the replica trick
$\overline{Z^k}=\int Du e^{-S_{\text{rep}}/\hbar}$ and integrating over
$\Pi$, after some manipulations \cite{giamarchi_columnar_variat},
one obtains the following replicated action $S_{\text{rep}} =
S_{\text{ph}} + S_{\text{dis}}$ with:
\begin{eqnarray}
&& S_{\text{ph}} = \int d^d x d\tau \frac{c}{2} \sum_a (\nabla_x
u_a)^2 + \frac{1}{v^2} (\partial_{\tau} u_a)^2  \nonumber \\
&& S_{\text{dis}}=-  \frac{1}{2 \hbar} \int d^d x d\tau d\tau'
\sum_{ab} R(u_a(x,\tau) - u_b(x,\tau')) \nonumber \\
&& R(u) = \rho_0^2 \sum_K \Delta_K \cos(K \cdot u)
\label{Srepliquee}
\end{eqnarray}
Here
$v=\sqrt{c/M}$ is the pure phonon velocity and
$\Delta_K = \int d^d x e^{i K \cdot x} \Delta(x)$ denote the
harmonics of the disorder correlator at the reciprocal lattice
vectors $K$, and $\rho_0 \sim a^{-2}$ the average areal density.

Given the complexity of the replicated action $S_{\text{rep}}$, it has
been proposed to study it within the Gaussian Variational Method (GVM)
\cite{giamarchi_vortex_long,mezard_variational_replica,giamarchi_columnar_variat}.  
It is implemented by choosing a trial gaussian action $S_0$
parametrized by a $k \times k$ matrix in replica space
$G^{-1}_{ab}(q,\omega_n)$:
\begin{eqnarray}\label{def_Sgauss}
&&S_0 = \frac{1}{2 \beta \hbar} \int_q \sum_{a,b}
G^{-1}_{ab}(q,\omega_n) u^a(q,\omega_n)u^b(-q,-\omega_n) \nonumber \\
&&G^{-1}_{ab}(q,\omega_n) = cq^2 \delta_{ab} -
\sigma_{ab}
\end{eqnarray}
which minimizes the variational free energy $F^{\text{var}} = F_0 +
\frac{1}{\beta \hbar} \langle S^{\text{rep}} - S_0  \rangle_{S_0}$,
where $F_0$ denotes the free energy computed with $S_0$.  
In the limit $k \to 0$, we denote $\tilde{G}(q,\omega_n) =
G_{aa}(q,\omega_n)$ and 
parametrize $G_{a \neq b}(q,\omega_n)$ by $G(q,u)$, where $0 < u
<1$, which is $\omega_n$ independent \cite{giamarchi_columnar_variat}.
Similarly we take $B_{ab}(\tau) =
\overline{\langle [u_a(x,\tau) - u_b(x,0)]^2\rangle}/m$ with
$\tilde{B}(\tau)$ and $B(u)$ which is $\tau$ independent. The best trial
Gaussian action (\ref{def_Sgauss}) is obtained by breaking 
the replica-symmetry (RSB) 
\cite{giamarchi_vortex_long}. A previous analysis
\cite{giamarchi_vortex_long} 
revealed indeed the 
existence of a breakpoint $u_c$ such that 
$\sigma(u) = \sigma(u_c)$ for $u \geq u_c$. In $d>2$, where there is a
full RSB solution, $\sigma(u)$ is a continuously  
varying function of $u$ for $u < u_c$. In $d = 2$, for the single
cosine model, there is
instead a (marginal) one step RSB solution such that $\sigma(u) = 0$ for
$u < u_c$.  

Using the variational approach, it has been shown in details
\cite{schehr_chalspe_long} that the specific heat is obtained from the
$T$-derivative of internal energy $\overline{\langle H \rangle}$ per
unit volume,
which, independently of the RSB scheme, can be written in terms of the
saddle point solution: 
\begin{widetext}
\begin{eqnarray}\label{H_quant_variat_simpl}
&&\overline{\langle H \rangle}
= \frac{1}{\beta} \sum_n \int_q
\frac{cq^2 + \Sigma + I(\omega_n)}{cq^2+\Sigma+M
  \omega_n^2+I(\omega_n)} 
+ \frac{1}{\hbar} \int_0^{\beta \hbar} d\tau (F(\tilde{B}(\tau)) -
  F(B)) - \int_0^{w_c} dw (F(B(w)) - F(B))
\end{eqnarray}
\end{widetext}
where $F(X) = \hat{V}(X) -
\frac{X}{2}\hat{V}'(X)$, $\hat{V}(X) = - \rho_0^2
\sum_K \Delta_K \exp(- X K^2/2)$, $w_c = \beta u_c$. In
(\ref{H_quant_variat_simpl}), the quantities entering this expression are
determined by the variational equations:
\begin{equation}
I(\omega_n) = \frac{2}{\hbar} \int_0^{\beta
\hbar} d\tau (1-\cos{(\omega_n \tau)}) (\hat{V}'(\tilde{B}(\tau)) -
\hat{V}'(B) ) \label{Eq_I} 
\end{equation}
\begin{eqnarray}
&&1 = - 4 \hat V''(B) J_2(\Sigma) \label{marginality} \\
&& J_n(x) = \int_q \frac{1}{(cq^2 + x)^n} \nonumber 
\end{eqnarray}
with the definitions
\begin{eqnarray}
&&B = \frac{2}{\beta} \sum_n \int_q \frac{1}{cq^2
+ M \omega_n^2 + \Sigma + I(\omega_n)}  \label{def_B} \\
&&\tilde{B}(\tau) = \frac{2}{\beta}\sum_n \int_q
  \frac{1-\cos{(\omega_n \tau)}}{cq^2 + M\omega_n^2 + \Sigma + I(\omega_n)}
 \label{def_b_tau} 
\end{eqnarray}
The breakpoint $w_c$, in $d>2$, is determined by:
\begin{eqnarray}
w_c \equiv w_c(\Sigma) = 4
\frac{({J}_2(\Sigma))^3}{{J}_3(\Sigma)} \hat{V}'''(B)\label{wc}
\end{eqnarray}
We finally quote the following useful relation, valid a full RSB
solution, obtained by combining Eq. (\ref{marginality}) and Eq. (\ref{wc}):
\begin{eqnarray}\label{rel_db_dsigma}
w_c \delta B + 2 {\cal J}_2(\Sigma)\delta \Sigma = 0
\end{eqnarray} 
where $\delta$ stands for an infinitesimal variation. 
The equation (\ref{marginality}) is the {\it marginality condition},
corresponding to the vanishing of the replicon eigenvalue, which
holds automatically in this problem for $d \geq 2$. As first noticed in
\cite{giamarchi_columnar_variat} and further investigated in
\cite{schehr_chalspe_quantique, schehr_chalspe_long}, the solution of
the variational equations can be organized in an
expansion in $\hbar$ keeping $\beta \hbar$ fixed. Expanding any
quantity $Q$ as $Q = \sum_{n=0}^{\infty} \hbar^n Q_n(\beta \hbar)$, it
was    
shown \cite{giamarchi_columnar_variat}
that this condition describes a {\it gapless} excitation spectrum
caracterized by the low frequency behavior of the self-energy
(\ref{Eq_I}) $I_0(\omega_n) \propto 
|\omega_n| + {\cal O}(\omega_n^2)$ leading to the analytic
continuation $I_0''(\omega) \propto \omega$ where $I(\omega_n \to -i \omega
+ 0^+)= I'(\omega) + i I''(\omega)$. Thus one expects a power law
behavior of the specific heat at low $T$. In the following, we will
compute analytically the low $T$ expansion of the internal energy
(\ref{H_quant_variat_simpl}).

\subsection{Quantum spherical $p$-spin glass model (Model II).}


We consider a quantum extension of the spherical $p$-spin glass
model as studied in \cite{cugliandolo_quantum_p_spin},
an interacting system of $N$ 
continuous spins $s_i$, $1<i<N$.  
This quantum extension consists in considering a continuous spin $s_i$
as an operator associated to a spatial coordinate and introducing its
conjugated momentum $\pi_i$ which satisfies standard commutation relations 
\begin{eqnarray}
[s_i,s_j] = [\pi_i,\pi_j] = 0 \quad , \quad [\pi_i, s_j] = - i\hbar
\delta_{ij} 
\end{eqnarray}
The quantum $p$-spin glass model is then described by the following
hamiltonian 
\begin{eqnarray}\label{def_pspin}
H[\overrightarrow{\pi},\overrightarrow{s},J] = \frac{\pi^2}{2M} +
\sum_{i_1 < .. < i_p}^N J_{i_1,..,i_p} s_{i_1}...s_{i_p}
\end{eqnarray}
where we denote $\pi^2 = \overrightarrow{\pi} \cdot
\overrightarrow{\pi}$, with $\overrightarrow{\pi} = (\pi_1,...,\pi_N)$
(similarly for $s^2$ and $\overrightarrow{s}$) and impose the
spherical constraint  
\begin{eqnarray}
\frac{1}{N} \sum_{i=1}^N \langle s_i^2 \rangle = 1
\end{eqnarray}
In (\ref{def_pspin}), the coupling constants $J_{i1,..,ip}$ are random
variables, independently distributed according to a gaussian distribution
of zero mean and variance 
\begin{eqnarray}\label{Def_Dis_p}
\overline{ J_{i_1,..,i_p}^2} = \frac{\tilde{J}^2 p !}{2 N^{p-1}}
\end{eqnarray}  
This model
(\ref{def_pspin}) is then studied \cite{cugliandolo_quantum_p_spin}
using the formalism of the quantum 
action in imaginary time 
(\ref{Def_Action}) together with the use of replicas to implement the
average over the disorder (\ref{Def_Dis_p}). After some manipulations,
one obtains, in the limit $N \to \infty$ a saddle point equation for
the order parameter
\begin{eqnarray}
Q_{ab}(\tau-\tau') = \frac{1}{N} \overline{\langle
  \overrightarrow{s}_a(\tau) \cdot \overrightarrow{s}_b(\tau')   \rangle} 
\end{eqnarray}
where $a,b$ are replica indices. 
In the limit $k \to 0$, one denotes $Q_{aa}(\tau) = q_d(\tau)$ and
parametrizes $Q_{a\neq b}(\tau)$ by $q(u)$ which is
$\tau$-independent. Following the authors of Ref. 
\cite{cugliandolo_quantum_p_spin}, we 
will work with dimensionless quantities, by redefining the imaginary
time $\hat\tau = \tilde J \tau/\hbar$ and Matsubara frequencies $\hat
\omega_n = \hbar \omega_n/\tilde J$ (in the following we will drop all
hats in order to simplify the notations). We also introduce the
parameter $\Gamma = \hbar^2/(M \tilde J) $, which measures the
strength of quantum fluctuations. The phase diagram of
(\ref{def_pspin}) in the 
$\Gamma - T$ plane was found \cite{cugliandolo_quantum_p_spin} to be
characterized by a line 
$\Gamma_c(T)$ separating a paramagnetic (PM), associated to a
diagonal matrix $Q_{ab}(\tau) = q_d(\tau) \delta_{ab}$, 
from a
spin-glass (SG) phase at low $T$, which we focus on here.   
The saddle point equations describing this SG phase 
is solved by a one step RSB ansatz \cite{cugliandolo_quantum_p_spin},
shown to 
be exact as in the classical case, such that $q(u) = 0$ for
$u<m$ and $q(u) =
q_{\text{EA}}$ for $u > m$, $m$ being the breakpoint. 
The internal energy, as a
function of the 
saddle point solution is given by \cite{cugliandolo_quantum_p_spin}:
\begin{eqnarray}\label{H_p_spin}
&&\overline{\langle H \rangle}  = \frac{z'}{2} + \frac{p}{4} \int_0^{\beta}
  d\tau (q_d^{p-1}(\tau) - q_{\text{EA}}^{p-1} ) \nonumber \\
&&- \frac{p+2}{4} \beta m q_{\text{EA}}^p - \frac{p+2}{4} \int_0^{\beta}
  d\tau (q_d^{p}(\tau) - q_{\text{EA}}^p)
\end{eqnarray}
with the saddle point equations
\begin{eqnarray}
\hspace*{-0.2cm}\tilde{\Sigma}(\omega_n) &=& \frac{p}{2} \int_0^{\beta} d\tau
  (1-\cos{(\omega_n \tau 
)})(q_d^{p-1}(\tau) - q_{\text{EA}}^{p-1} ) \label{Eq_Sigma} \\
\hspace*{-0.2cm} z' &=& \frac{p}{2} \beta m q_{\text{EA}}^{p-1}
\frac{1+x_p}{x_p} 
\label{eq_z} 
\end{eqnarray}
and the definitions
\begin{eqnarray}
q_{\text{EA}} = 1 - &&\frac{1}{\beta} \sum_n \frac{1}{
    \frac{\omega_n^2}{\Gamma}  
+ z' + \tilde{\Sigma}(\omega_n) 
  } \\
q_d(\tau) - q_{\text{EA}} = &&\frac{1}{\beta} \sum_n
\frac{\cos{\omega_n \tau}}{\frac{\omega_n^2}{\Gamma} 
+ z' + \tilde{\Sigma}(\omega_n)} \label{Green_pspin}
\end{eqnarray}
The breakpoint is determined by 
\begin{eqnarray}
 \beta m = x_p \sqrt{\frac{2}{p(x_p+1)}} q_{\text{EA}}^{-p/2} \label{uc_p}
\end{eqnarray}
Combining (\ref{eq_z}) and (\ref{uc_p}), one obtains the following
useful identity
\begin{eqnarray}\label{1surzp}
\frac{1}{z'^2} = \frac{2}{p(1+x_p)} q_{\text{EA}}^{2-p} 
\end{eqnarray}
As it was noticed in other one step RSB solution
\cite{georges_mf_quantum_spinglass,giamarchi_columnar_variat},  
one obtains a one parameter family of solutions, indexed by
$x_p$ (or equivalently by the breakpoint $m$). There are then two
different ways to determine $m$.  
In the statics, $m$ is
usually determined by minimizing the free energy :
this is the so-called {\it equilibrium} criterion. The excitation
spectrum of the equilibrium SG state is gaped, yielding a
specific heat wich vanishes {\it exponentially} at low $T$.   
Alternatively, $m$ is determined by imposing the vanishing of the
replicon eigenvalue, which leads to the so called {\it marginality}
condition \cite{cugliandolo_quantum_p_spin}: 
\begin{eqnarray}
&&x_p = p-2 \label{marginality_p}
\end{eqnarray}
One can
show \cite{cugliandolo_quantum_p_spin}, using a Keldysh mean field
approach and performing analytical 
continuation to imaginary time, 
that the marginality condition (\ref{marginality_p}) gives 
indeed the correct solution from the dynamical point of view,
{\it i.e.}, if one considers, in an infinite system, the
large time limit where time translational invariance 
and equilibrium fluctuation dissipation theorem hold. Moreover, 
this marginal value of $x_p$ was found to be the only one compatible
with a {\it gapless} excitation spectrum
\cite{cugliandolo_quantum_p_spin}. In the $T=0$ limit, it was 
indeed shown that, in the low frequency limit
$\tilde{\Sigma}(\omega_n)  \propto |\omega_n| + {\cal
  O}(\omega_n^2)$. Therefore, one expects
that the specific heat of the marginally stable SG state vanishes as a
power law. In the following, we show how to extract analytically the
low $T$ behavior of the marginally stable SG state. 


\subsection{Quantum SU(N) spin-glass (Model III).}


We consider the Heisenberg quantum spin-glass, defined by the
following hamiltonian\cite{sachdev_ye_sg,georges_mf_quantum_spinglass}  
\begin{eqnarray}\label{suN}
H = \frac{1}{N {\cal N}} \sum_{ij} J_{ij} \overrightarrow{S}_i \cdot
\overrightarrow{S}_j 
\end{eqnarray}
where the original spin symmetry group $SU(2)$ is extended to 
$SU(N)$ \cite{sachdev_ye_sg} and the large $N$ limit is taken. 
These ${\cal N}$ spins occupy the sites of a fully
connected lattice. In (\ref{suN}), the coupling $J_{ij}$ are 
random variables, independently distributed according a gaussian
distribution of zero mean and variance 
\begin{eqnarray}\label{Def_Dis_suN}
\overline{J_{ij}^2} = J^2 
\end{eqnarray}
Using the imaginary time path-integral formalism (we will set $\hbar =
1$ from the beginning for this model), together with
replicas to implement the average over the disorder
(\ref{Def_Dis_suN}), the model is mapped, in the
infinite range limit, onto a self-consistent
single site problem described by the action \cite{bray_quantum_sg}:
\begin{eqnarray}\label{S_eff}
S_{\text{eff}} = S_B - \frac{J^2}{2 N} \int_0^{\beta} d\tau d\tau' {\cal
Q}_{ab}(\tau-\tau') \overrightarrow{S}_a(\tau) \cdot
\overrightarrow{S}_b(\tau') 
\end{eqnarray}
where $S_B$ is the Berry phase imposing the spin commutation relations
\cite{sachdev_ye_sg},
together with the 
self-consistent equation:
\begin{eqnarray}\label{start_self_consist}
{\cal Q}_{ab}(\tau-\tau') = \frac{1}{N^2} \langle
 \overrightarrow{S}_a(\tau) \cdot 
\overrightarrow{S}_b(\tau')  \rangle_{S_{\text{eff}}} 
\end{eqnarray}
where $\langle...\rangle_{S_{\text{eff}}}$ stands for an average
computed with the action $S_{\text{eff}}$.
Using a bosonic representation of the spin operator
$S$ in terms of Schwinger bosons,
$S_{\alpha \beta} = b^{\dagger}_{\alpha}b_{\beta} - S \delta_{\alpha
  \beta}$ with the constraint $\sum_\alpha b^{\dagger}_\alpha b_\alpha =
  S N$, this model (\ref{suN}) can be described 
analytically in the limit $N \to \infty$ which then constitutes a mean
field theory of the fully connected model (\ref{suN}) where the spins
have the symmetry $SU(2)$. In this limit, the original
self consistent equation (\ref{start_self_consist}) reduces to an
equation for the boson Green's function ${\cal G}^{ab}(\tau)
\equiv - \overline{\langle T b^a(\tau) b^{\dagger b}(0) \rangle}$. 
In the limit $k \to 0$, one parametrizes ${\cal G}^{aa}(\tau)$
by $\tilde{\cal G}(\tau) - \tilde{g}$, such that $\lim_{\tau \to \infty}
\tilde{\cal G}(\tau) = 0$ at $T=0$ and, ${\cal G}^{a\neq b}(\tau)$ by
$-g(u)$, which is $\tau$-independent. The phase diagram of (\ref{suN})
in the $T-S$ plane has been established in the large $N$ limit
\cite{georges_mf_quantum_spinglass}. A line $S_c(T)$ 
separates a paramagnetic phase, described
by a diagonal matrix in replica space ${\cal G}^{ab}(\tau) = \delta_{ab}
\tilde{\cal G}(\tau)$, and where several crossovers were found to
occur in the quantum regime \cite{georges_mf_quantum_spinglass}, from
a spin glass phase, which we focus on here. In this SG phase, the
saddle point equations are solved by a one step RSB ansatz, such that
$g(u)=0$ for $u<x$ and $g(u) = g$ for $u>x$, $x$ being the breakpoint
and $\tilde{g} = g$. The starting point of our computation of 
the specific heat is the expression for the internal
energy per unit volume \cite{georges_mf_quantum_spinglass}:
\begin{eqnarray}\label{H_sun}
&&\overline{\langle H \rangle} = -\frac{J^2}{2} \int_0^{\beta} d\tau
(\tilde{\cal G}(\tau) -g )^2(\tilde{\cal G}(-\tau) -g)^2 \nonumber \\
&&- \frac{J^2}{2} \beta(x-1)g^4
\end{eqnarray} 
in terms of the saddle point solution
\begin{eqnarray}\label{eqvar_sun}
&&\hat{\Sigma}(i\nu_n) = J^2 \int_0^{\beta}
  d\tau (e^{-i\nu_n \tau} - 1) \label{Eq_I_sun} \\
&&\times \left((\tilde{\cal G}(\tau) -
  g)^2(\tilde{\cal G}(-\tau) - 
g) + g^3\right)\nonumber \\
&&\beta x = \frac{1}{Jg^2}\left(\frac{1}{\Theta} - \Theta  \right)
  \label{uc_sun}  
\end{eqnarray}
with the definitions
\begin{eqnarray}\label{def_Green_suN}
&&g = S + \tilde{\cal G}(\tau = 0^{-}) \nonumber \\
&&\tilde{\cal G}(i\nu_n) = \frac{1}{i\nu_n - \frac{Jg}{\Theta} -
    \hat{\Sigma}(i\nu_n)} 
\end{eqnarray}
where $\nu_n$ is a bosonic Matsubara frequency. Similarly to the
spherical $p$-spin model (\ref{Eq_Sigma}-\ref{uc_p}), one obtains a
one parameter family 
of solutions, parametrized by $\Theta$, or equivalently by the
breakpoint $x$. Here also, if one chooses the {\it equilibrium}
criterion, the excitation spectrum is gapped. Instead, if one imposes
the vanishing of the replicon eigenvalue, one obtains the {\it marginality}
condition \cite{georges_mf_quantum_spinglass}:   
\begin{eqnarray}\label{margin_sun}
\Theta = \Theta_R = \frac{1}{\sqrt{3}}
\end{eqnarray}
Using an expansion in $1/S$ -- similar to the semi-classical expansion
for the elastic manifold \cite{giamarchi_columnar_variat} --, it has
been shown explicitly \cite{georges_mf_quantum_spinglass} that the
marginality condition (\ref{margin_sun}) is the only one compatible
with a gapless excitation spectrum, such that
$\hat{\Sigma}''(\omega) \propto \omega + {\cal
  O}(\omega^2)$,
where $\hat{\Sigma}(\nu_n \to -i\omega + 0^+)  =
\hat{\Sigma}(\omega) + i\hat{\Sigma}''(\omega)$. 
Although the
connexion between this Matsubara formalism using the
marginality condition (\ref{margin_sun}) and the true hamiltonian
dynamics in real time at the Mean Field level has not yet been 
established for the present case, the study of similar one step RSB
solutions for which this connexion has been done
\cite{cugliandolo_quantum_p_spin, cugliandolo_keldysh_elastic},
suggests that (\ref{margin_sun}) gives indeed the correct solution from the
dynamical point of view. On physical grounds, 
in the present case of the ordered phase of a 
quantum SG with continuous symmetry, this choice seems also
natural as one indeed expects a gapless excitation spectrum
\cite{georges_mf_quantum_spinglass}.    

Although the
specific heat in the equilibrium SG state vanishes
exponentially at low $T$, we focus here on the low $T$ behavior of
$C_v(T)$ in the marginal state (\ref{margin_sun}), which is instead
expected to vanish as a power law \cite{georges_mf_quantum_spinglass}.

\section{LOW TEMPERATURE ANALYSIS.}

In this section, we compute the low temperature expansion of the
internal energy for the different models presented before, from which
we directly obtain the specific heat 
$C_v(T)$. To do so, we start by deriving some general identities,
which form the background of our analysis. The finite temperature
behavior of the internal energy requires the low temperature expansion
of the saddle point equations, which we then present in details. For
that purpose, we will use the notation, for any quantity of interest
$Q$, $Q = \sum_{n =0}^{\infty} T^{n} Q^{(n)}$. We finally turn
to the behavior of $C_v(T)$ in the last paragraph of this section.

\subsection{General properties.}

We first focus on the two first problems
evocated here (\ref{Hsys},\ref{def_pspin}), which show, formally, a strong
similarity. In particular, at variance with Model III
(\ref{suN}), these two 
systems exhibit a two point
Green's function (\ref{def_b_tau}, \ref{Green_pspin}) wich is invariant
under the transformation $\omega_n \to - \omega_n$. We focus on the
low $T$ behavior of integrals over imaginary time $\tau$ which enter
both the variational equation and the computation of the internal
energy. We will use the notations of the disordered elastic
hamiltonian (\ref{Eq_I}-\ref{wc}), the transposition to the $p$-spin
model being straightforward.    

We first suppose that $\Sigma$ and $I(\omega_n)$, as a function of
$\omega_n$ are independent of $T$, {\it i.e.} $\Sigma^{(n)} = 0$,
$\forall n > 0$ and similarly for $I(\omega_n)$ with $I(\omega_n) \sim
|\omega_n| + {\cal O}(\omega_n^2)$. Then, as shown in
Appendix \ref{app_low_T_gen_BG}, one has the low temperature
expansion,for any function ${\cal H}(X)$ that can be expanded as a 
power serie around $0$, ${\cal H}(X) = \sum_{k=0}^{\infty} a_k X^k$ : 
\begin{equation}\label{Property_1_bg}
\int_0^{\beta \hbar} d\tau \cos{(\omega_n \tau)} ({\cal H}(\tilde{B}(\tau)) -
 {\cal H}(B)) 
 = C^{\text{st}} + {\cal O}(T^4)
\end{equation}
where here, and in the following, $C^{\text{st}}$ stands for a (generic)
quantity independent of $T$ (it may however depend on the Matsubara
frequency). Indeed,
the gapless structure of the spectral function suggests that only even
powers of $T$ should enter this expansion (\ref{Property_1_bg}). But
as shown in details in the Appendix \ref{app_low_T_gen_BG}, one can
check explicitly that the term $\propto T^2$ {\it vanishes}. This
property (\ref{Property_1_bg}) is rather independent of the structure
of the variational equations, only requiring $I''(\omega) \propto
\omega$ at low frequency.    

For the other model discussed here (\ref{suN}), one could not show a
so general 
statement (\ref{Property_1_bg}) involving any function ${\cal H}$, due to the
absence of the symmetry $i \nu_n \to 
-i\nu_n$ of the Green's function $\tilde{\cal G}(i \nu_n)$
(\ref{def_Green_suN}) 
, which renders the calculations more subtle. 
Nevertheless, with the same assumptions as above that the complete
self-energy $Jg/\theta + \hat{\Sigma}(i\nu_n)$ (\ref{def_Green_suN}) is
independent of $T$, with $\hat{\Sigma}(i\nu_n) \propto |\nu_n| + {\cal
O}(\nu_n^2)$,   
for the particular integrals involved here, one can show (see Appendix
\ref{app_low_T_gen_suN} for more details):
\begin{eqnarray}\label{Property_1_suN}
&&\hspace*{-0.5cm}\int_0^{\beta} d\tau ((\tilde{\cal G}(\tau) -
g)^2(\tilde{\cal G}(-\tau) - g)^2 - g^4) = C^{\text{st}} + {\cal O}(T^4) \\
&&\hspace*{-0.5cm}\int_0^{\beta}d\tau e^{-i\nu_n \tau} ((\tilde{\cal G}(\tau)
-g)^2(\tilde{\cal G}(-\tau) -g) + g^3) = C^{\text{st}} + {\cal O}(T^4)
\nonumber  
\end{eqnarray} 
 
These properties are very useful tools to investigate the low $T$
behaviors of physical quantities in these models.

\subsection{Variational equations.}

We now turn to the low $T$ expansion of the saddle point equations for
the three different models.

\subsubsection{Model I.}

Most of the properties presented here, and their extension to the spin-glass
models (\ref{def_pspin}, \ref{suN}), have been suggested by the
expansion in powers 
of $\hbar$, 
keeping $\beta \hbar$ fixed. We will shortly remind here the main features of
the semi classical expansion of the variational equations (\ref{Eq_I},
\ref{marginality}). We use 
the notation, for any 
quantity $Q = \sum_{k=0}^{\infty} \hbar^k Q_k(\beta \hbar)$. Although
at lowest 
order, the complete solution of the variational equations
\cite{giamarchi_columnar_variat} shows that
$\Sigma_0$ and $I_0(\omega_n)$ are independent of $T$, $\Sigma_1$ and
$I_1(\omega_n)$ become $T$ dependent at the next order
\cite{schehr_chalspe_long}. At low temperature the following structure
was {\it explicitly} obtained
\begin{eqnarray}
&&\Sigma_1 = \Sigma_1^{(0)} + \left(\frac{T}{\hbar}\right)^2
\Sigma_1^{(2)} + O((T/\hbar)^4) \label{sigma_1}\\
&&I_1(\omega_n) = I_1^{(0)}(\omega_n) +  \left(\frac{T}{\hbar}\right)^2
I_1^{(2)}(\omega_n) +  O((T/\hbar)^4) \nonumber \\
&&I_1^{(2)}(\omega_n) = - (1-\delta_{n,0}) \Sigma_1^{(2)}
\label{peculiar_hbar1} 
\end{eqnarray}
such that the finite temperature corrections of $\Sigma_1 +
I_1(\omega_n)$ are $\propto T^4$ for $\omega_n \neq 0$. It was also
noticed that the peculiar term (\ref{peculiar_hbar1}), once
inserted in $B$ (\ref{def_B}), generates odd powers of $T$, the lowest
one being 
$T^3$, in the low $T$ expansion at higher order in the expansion in
$\hbar$. We extend here these 
properties (\ref{sigma_1}, \ref{peculiar_hbar1}) independently on the  
semi-classical approximation. We show indeed that the following expansion,
up to order ${\cal O}(T^4)$:
\begin{eqnarray}
&&\hspace*{-0.5cm}\Sigma = \Sigma^{(0)} + T^2 \Sigma^{(2)} + T^3
  \Sigma^{(3)}  \label{devel_variat_BG} \\
&&\hspace*{-0.5cm}I(\omega_n) = I^{(0)}(\omega_n) -
  (1-\delta_{n,0})(T^2 \Sigma^{(2)} 
  + T^3 \Sigma^{(3)})  \nonumber
\end{eqnarray}
with $I^{(0)}(\omega_n) \propto |\omega_n| + {\cal O}(\omega_n^2)$, 
is a consistent solution of the variational equations
(\ref{Eq_I},\ref{marginality}). To do so, we compute the low $T$
expansion of the r.h.s of the equation for $I(\omega_n)$ given the
forms (\ref{devel_variat_BG}). We thus need an extension of the
general property (\ref{Property_1_bg}), when $\Sigma$ and
$I(\omega_n)$ have the 
form given in Eq. (\ref{devel_variat_BG}). As shown in Appendix
\ref{app_low_T_gen_BG} , the 
low $T$ behavior of such integrals over $\tau$ (\ref{Property_1_bg})
are in that case given by
\begin{eqnarray}\label{Property_BG_2}
&&\int_0^{\beta \hbar} d\tau \cos{(\omega_n \tau)} ({\cal
 H}(\tilde{B}(\tau)) - {\cal H}(B)) 
 = C^{\text{st}} \\
&&+ 2 \hbar \delta_{n,0} (T^2 \Sigma^{(2)} + T^3
 \Sigma^{(3)}) J_2(\Sigma^{(0)}) {\cal H}'(B^{(0)}) + {\cal O}(T^4) \nonumber
\end{eqnarray}
If one applies this general formula (\ref{Property_BG_2}) with ${\cal H} =
2\hat{V}'(B)$ to the r.h.s of Eq. (\ref{Eq_I}), one obtains up to
order ${\cal O}(T^4)$:
\begin{eqnarray}\label{consistency}
&&\hspace*{-0.5cm}\frac{2}{\hbar} \int_0^{\beta
\hbar} d\tau (1-\cos{(\omega_n \tau)}) (\hat{V}'(\tilde{B}(\tau)) -
\hat{V}'(B) ) = C^{\text{st}} \\
&&\hspace*{-0.5cm} + 4 (1-\delta_{n,0})(T^2 \Sigma^{(2)} + T^3
 \Sigma^{(3)}) J_2(\Sigma^{(0)})J_2(\Sigma^{(0)}) \hat{V}''(B^{(0)})
 \nonumber \\
&&\hspace*{-0.5cm} = C^{\text{st}} -  (1-\delta_{n,0})(T^2
 \Sigma^{(2)} + T^3 \Sigma^{(3)})  
\end{eqnarray}
where, in the last line, we have used the marginality condition
(\ref{marginality}) at $T=0$. This relation (\ref{consistency}) thus
shows explicitly the consistency of the low $T$ expansion
(\ref{devel_variat_BG}) proposed for  
the exact solution of the variational equations. Importantly, although the
general property shown above (\ref{Property_BG_2}) holds
independently of the saddle point 
equation, a solution such as (\ref{devel_variat_BG}) is consistent
{\it provided the marginality condition holds}.

\subsubsection{Model II.}

Inspired by the previous analysis, we show that a consistent solution
of the variational equations for the quantum $p$-spin model
(\ref{Eq_Sigma}, \ref{eq_z}) is given, up
to order ${\cal O}(T^4)$ by:
\begin{eqnarray}\label{devel_variat_pspin}
&&z' = z'^{(0)} + T^2 z'^{(2)} + T^3 z'^{(3)}  \\
&&\tilde{\Sigma}(\omega_n) =
  \tilde{\Sigma}^{(0)}(\omega_n)  
   - (1-\delta_{n,0}) (T^2 z'^{(2)} + T^3 z'^{(3)}) \nonumber 
\end{eqnarray}
with $\tilde{\Sigma}^{(0)}(\omega_n) \propto |\omega_n| + {\cal
O}(\omega_n^2)$.  
This is shown by using the general low $T$ expansion, an extension of
(\ref{Property_BG_2}) to the present case
(\ref{devel_variat_pspin}):
\begin{eqnarray}\label{Property_pspin_2}
&&\int_0^{\beta \hbar} d\tau \cos{(\omega_n \tau)} ({\cal H}(q_d(\tau)) -
 {\cal H}(q_{\text{EA}})) 
 = C^{\text{st}}  \nonumber \\
&&- \delta_{n,0} (T^2 z'^{(2)} + T^3
 z'^{(3)}) \frac{{\cal H}'(q_{\text{EA}}^{(0)})}{(z'^{(0)})^2} + {\cal
 O}(T^4)
\end{eqnarray}
Thus applying (\ref{Property_pspin_2}) with ${\cal H}(X) = (p/2)
X^{p-1}$ yields 
the low $T$ expansion of the r.h.s of the variational
equation (\ref{Eq_Sigma}) up to order ${\cal O}(T^4)$
\begin{eqnarray}\label{Devel_inter_pspin}
&&\frac{p}{2} \int_0^{\beta} d\tau (1-\cos{(\omega_n \tau
)})(q_d^{p-1}(\tau) - q_{\text{EA}}^{p-1} ) = C^{\text{st}} \nonumber \\
&& - \frac{p(p-1)}{2}(1-\delta_{n,0})(T^2 z'^{(2)} + T^3
 z'^{(3)}) \frac{(q_{\text{EA}}^{(0)})^{p-2}}{(z'^{(0)})^2} \nonumber \\
&& = C^{\text{st}} - \frac{p-1}{1+x_p}(1-\delta_{n,0}) (T^2 z'^{(2)} + T^3
 z'^{(3)})
\end{eqnarray}
where we have used, in the last line, the identity (\ref{1surzp}).
Thus, one sees
on (\ref{Devel_inter_pspin}) that the expression given in
Eq. (\ref{devel_variat_pspin}) is a 
consistent solution of (\ref{Eq_Sigma}) 
provided $x_p = p-2$, {\it i.e.} the solution is marginally stable
(\ref{marginality_p}). 

\subsubsection{Model III.}

For this model, we show that, similarly to the two previous ones, a
consistent solution of the variational equations (\ref{eqvar_sun}) 
is given, up to order ${\cal O}(T^4)$ by
\begin{eqnarray}\label{Devel_variat_sun}
&&g = g^{(0)} + T^2 g^{(2)} + T^3 g^{(3)} \\
&&\hat{\Sigma}(i \nu_n)  =
  \hat{\Sigma}^{(0)}(i\nu_n) + \frac{J}{\Theta}(\delta_{n,0}-1) (T^2
  g^{(2)} + T^3 g^{(3)}) \nonumber 
\end{eqnarray}
with $\hat{\Sigma}^{(0)}(i\nu_n) \propto |\nu_n| + {\cal O}(\nu_n^2)$.
To show the consistency of this solution (\ref{Devel_variat_sun}), we
perform the low temperature expansion of the r.h.s of the equation for
$\hat{\Sigma}(i\nu_n)$ (\ref{eqvar_sun}) given
(\ref{Devel_variat_sun}). One obtains up to order 
${\cal O}(T^4)$:
\begin{eqnarray}\label{Devel_inter_sun}
&&J^2\int_0^{\beta}d\tau (e^{-i\nu_n \tau}-1) [(\tilde{\cal G}(\tau)
-g)^2(\tilde{\cal G}(-\tau) -g) + g^3]  \nonumber \\
&&= C^{\text{st}} + (\delta_{n,0}-1)3\Theta J
\left( T^2 g^{(2)}   +
T^3 g^{(3)}   \right)
\end{eqnarray} 
Again, this expansion (\ref{Devel_inter_sun}) shows that the structure
exhibited 
in (\ref{Devel_variat_sun}) is a consistent solution of the
variational equation (\ref{eqvar_sun}) provided $3 \Theta = 1/\Theta$,
{\it i.e.} $\Theta = \Theta_R = 1/\sqrt{3}$, the marginality condition
(\ref{margin_sun}). 

\subsection{Specific heat : low temperature expansion.}

We now turn to the computation of the low temperature behavior of the
specific heat. 

\subsubsection{Model I.}

Our starting point is the expression for the variational internal
energy $\overline{\langle H \rangle}$ given in
(\ref{H_quant_variat_simpl}). We first analyse the low temperature
behavior of the first term in (\ref{H_quant_variat_simpl}), namely the
sum over Matsubara frequencies. Importantly, we notice that, in this
sum, the contribution of the mode $\omega_n=0$ is independent of the
peculiar structure of $I(\omega_n)$ exhibited in
(\ref{peculiar_hbar1}) : this allows to avoid ambiguities
coming from the analytic continuation of such a term $\propto
(1-\delta_{n,0})$ in (\ref{peculiar_hbar1}). We can thus safely transform
this discrete sum in an integral: 
\begin{eqnarray}
&&\frac{1}{\beta} \sum_n \int_q
\frac{cq^2 + \Sigma + I(\omega_n)}{cq^2+\Sigma+M
  \omega_n^2+  I(\omega_n)} \\
&& = 
\int_{-\infty}^{+\infty} \frac{d\omega}{\pi} \hbar \omega
\rho_{\text{\tiny DOS}}(\omega) f_B(\omega) \label{H1_integral} \\
&&\rho_{\text{\tiny DOS}}(\omega) = \int_q \frac{-
  \frac{c}{v^2} \omega
  I''(\omega)}{(cq^2 - 
  \frac{c}{v^2} \omega^2 + \Sigma + \hat I'(\omega)^2 +
  (I''(\omega))^2)} \nonumber
\end{eqnarray}
where $\rho_{\text{\tiny DOS}}(\omega)$ is the density of states. Using
that $\rho_{\text{\tiny DOS}}(\omega) = 
\rho_{\text{\tiny DOS}}^{(0)}(\omega) + T^4 
\rho_{\text{\tiny DOS}}^{(4)}(\omega)$ (\ref{devel_variat_BG}), together with the low
frequency behavior $\rho_{\text{\tiny DOS}}^{(0)} \propto \omega^2$, one obtains
straighforwardly
\begin{equation}\label{H_low_T_first}
\frac{1}{\beta} \sum_n \int_q
\frac{cq^2 + \Sigma + \hat I(\omega_n)}{cq^2+\Sigma+M
  \omega_n^2+\hat I(\omega_n)} = C^{\text{st}} + {\cal O}(T^4) 
\end{equation}
One performs the low $T$ expansion of the 
second term in (\ref{H_quant_variat_simpl}), the integral over
  $\tau$, using the property (\ref{Property_BG_2}) which holds for the
  solution of the variational equations we have found (\ref{devel_variat_BG}): 
\begin{eqnarray}\label{H_low_T_second}
&&\int_0^{\beta \hbar} d\tau (F(\tilde{B}(\tau)) - F(B))
 = C^{\text{st}} \\
&&+ 2 (T^2 \Sigma^{(2)} + T^3
 \Sigma^{(3)}) J_2(\Sigma^{(0)}) F'(B^{(0)}) + {\cal O}(T^4) \nonumber
\end{eqnarray}
One finally tackles the analysis of the last term in
(\ref{H_quant_variat_simpl}), the integral over $w$,  
by using the property that $B(w)$ is independent of $T$ for $w <
w_c$ \cite{schehr_chalspe_long}.
\begin{eqnarray}\label{H_low_T_third}
&&\int_0^{w_c} dw (F(B(w)) - F(B)) = C^{\text{st}} \nonumber \\
&&+ w_c^{(0)} (T^2
B^{(2)} + T^3B^{(3)})F'(B^{(0)}) + {\cal O}(T^4)
\end{eqnarray} 
This property (\ref{H_low_T_third}) is then obvious for a full RSB
solution where $B(w_c^-) = B$ and was shown \cite{schehr_chalspe_long}
to hold also for the 
marginal one step RSB solution in $d=2$. 

Using the consequence of the marginality condition (\ref{rel_db_dsigma}),
one thus sees clearly that the quadratic and cubic contributions to
this low $T$ behavior {\it cancel} between Eq. (\ref{H_low_T_second}) and
Eq. (\ref{H_low_T_third}). And this yields
\begin{eqnarray}\label{H_low_T_final}
\overline{\langle H \rangle} = C^{\text{st}} + {\cal O}(T^4)
\end{eqnarray} 
This results in the
low temperature behavior of the specific heat:
\begin{eqnarray}\label{Cv_BG}
C_v(T) \propto T^3 + {\cal O}(T^4)
\end{eqnarray}
Although the coefficient of this cubic term is very hard to extract by
the method presented here, it has been explictly
computed at the lowest order in the aforementioned semi-classical
expansion and it was found to be non zero \cite{schehr_chalspe_quantique}.

\subsubsection{Model II.}

We analyse the low $T$ behavior of the internal energy within the
variational method (\ref{H_p_spin}) of the quantum $p$-spin model. We
treat separately the two lines of (\ref{H_p_spin}). Using
(\ref{Property_pspin_2}), one obtains
\begin{eqnarray}\label{H_p_first_part_inter}
&&\frac{p}{4} \int_0^{\beta}
  d\tau (q_d^{p-1}(\tau) - q_{\text{EA}}^{p-1} ) =  C^{\text{st}} \\
&& - (T^2 z'^{(2)} +
  T^3z'^{(3)})\frac{p(p-1)}{4}\frac{
  (q_{\text{EA}}^{(0)})^{p-2}  }{(z'^{(0)})^2}  + {\cal O}(T^4) \nonumber
\end{eqnarray}
Using this expansion and the identity (\ref{1surzp})
one obtains that the quadratic
and cubic terms in the low $T$ expansion of the first line in
(\ref{H_p_spin}) cancel provided $x_p = p-2$ (\ref{marginality_p}):
\begin{equation}\label{H_p_first_part}
\frac{z'}{2} + \frac{p}{4} \int_0^{\beta}
  d\tau (q_d^{p-1}(\tau) - q_{\text{EA}}^{p-1} ) =  C^{\text{st}} +
  {\cal O}(T^4) 
\end{equation}
We now focus on the second line of (\ref{H_p_spin}). Using
(\ref{Property_pspin_2}), one has immediately
\begin{eqnarray}\label{H_p_scd_part_inter_1}
&&- \frac{p+2}{4} \int_0^{\beta}
  d\tau (q_d^{p}(\tau) - q_{\text{EA}}^p) = C^{\text{st}} \\
&&+ (T^2 z'^{(2)} + T^3
 z'^{(3)})
  \frac{p(p+2)}{4}\frac{ (q_{\text{EA}}^{(0)})^{p-1}    }{(z'^{(0)})^2}    
  + {\cal O}(T^4) \nonumber
\end{eqnarray}
Finally combining (\ref{uc_p}) with
(\ref{1surzp}), one obtains the
expansion of the 
remaining term in (\ref{H_p_spin}) up to order ${\cal O}(T^4)$:
\begin{equation}\label{H_p_scd_part_inter_2}
-\frac{p+2}{4}\beta m q_{\text{EA}}^p = - \frac{p(p+2)}{4}
 \frac{z'^{(2)}T^2+z'^{(3)}T^3   }{(z'^{(0)})^2}
 (q_{\text{EA}}^{(0)})^{p-1} 
\end{equation} 
Collecting the different contributions to $\overline{\langle H
  \rangle}$
  (\ref{H_p_first_part_inter},
  \ref{H_p_scd_part_inter_1},\ref{H_p_scd_part_inter_2}), one obtains
\begin{eqnarray}
\overline{\langle H \rangle} = C^{\text{st}} + {\cal O}(T^4)
\end{eqnarray} 
 which leads to the low temperature behavior of the specific heat
\begin{eqnarray}\label{Cv_p}
C_v(T) \propto T^3 + {\cal O}(T^4)
\end{eqnarray}

\subsubsection{Model III.}

The low $T$ expansion of the internal energy of the Heisenberg
spin-glass model (\ref{H_sun}) is performed using the relation derived
in Appendix \ref{app_low_T_gen_suN} which, given the solution of the
variational equations we 
have found, yield
\begin{eqnarray}\label{H_sun_devel1}
&&-\frac{J^2}{2}\int_0^{\beta} d\tau [(\tilde{\cal G}(\tau) -
g)^2(\tilde{\cal G}(-\tau) - g)^2 - g^4] = C^{\text{st}}  \nonumber \\
&&+ 2{J \Theta}g^{(0)} \left(g^{(2)}T^2   +
g^{(3)}  T^3  \right) + {\cal O}(T^4)
\end{eqnarray} 
The expansion of the last term in (\ref{H_sun}) is straighforwardly computed
using the relation (\ref{uc_sun})
\begin{eqnarray}\label{H_sun_devel2}
&&-\frac{J^2}{2} \beta x g^4 = C^{\text{st}} \nonumber \\
&&- J g^{(0)}\left(\frac{1}{\Theta} - \Theta \right)
 (g^{(2)}T^2+g^{(3)}T^3) + {\cal O}(T^4)
\end{eqnarray}
Thus combining Eq. (\ref{H_sun_devel1}) and
Eq. (\ref{H_sun_devel2}), one sees that the quadratic and cubic
terms come with a prefactor $3\Theta -1/\Theta$, which thus vanishes
for the marginally stable solution, corresponding to
$\Theta=\Theta_R=1/\sqrt{3}$. This yields the low temperature behavior
of the specific heat
\begin{eqnarray}\label{Cv_sun}
C_v(T) \propto T^3 + {\cal O}(T^4)
\end{eqnarray}
For this model too, the amplitude of the cubic term has been computed
in a $1/S$, semi-classical, expansion, and found to be non-zero
\cite{georges_mf_quantum_spinglass}.   

\section{CONCLUSION.}

To sum up, we have computed the low temperature specific heat of a
rather wide class of quantum disordered systems with continuous degrees
of freedom, using a Mean Field
approximation, including disordered
elastic systems in $d \geq 2$, the spherical $p$-spin-glass and
the Heisenberg spin-glass. For all these models, we have obtained that
the Mean Field approximation yields the low $T$ behavior $C_v(T) \propto T^3$
(\ref{Cv_BG},
\ref{Cv_p},\ref{Cv_sun}). For the Heisenberg spin-glass model, the
cancellation of the linear 
term in $C_v(T)$ obtained here is in agreement with the numerical
solution of the 
saddle point equation of Ref. \cite{camjayi_sun_sg_num}. And
the non trivial structure of the saddle point solution
elucidated here (\ref{Devel_variat_sun}) could help to clarify
numerically the status of the quadratic contribution to $C_v(T)$
obtained in Ref. \cite{camjayi_sun_sg_num}.

This Mean Field result $C_v(T) \propto T^3$ is at variance with the
linear behavior commonly expected from two-level systems 
\cite{anderson_twolevels}. As we have shown, the cancellation of the
linear and quadratic 
contributions to $C_v(T)$ strongly relies upon the marginality
condition. And it is worthwhile to notice that the physical picture
associated to this marginal stability criterion, which enforces the
(quantum) dynamics along the flat 
directions of the free energy landscape, seems qualitatively different
from the argument steming from TLS
\cite{anderson_twolevels}.  

For the case of manifolds, our result,
within the Gaussian Variational Approximation also applies to
non periodic elastic structures, {\it e.g.} domain walls, when they can be
solved by a full RSB ansatz (or its limiting case of a marginal one
step RSB). And although this Mean Field approach is
always an approximation for the periodic case, it becomes exact for
the non periodic one, in the limit where the number of components of
the displacement field becomes infinite, as it is  
for the spherical $p$-spin-glass model
(\ref{def_pspin}), in the limit $N \to \infty$, or
for the Heisenberg spin-glass model (\ref{suN}) when both $N,{\cal
  N} \to \infty$. An outstanding question remains to know whether and
  how this
result is modified away from mean field, which clearly deserves further
numerical and analytical investigations.

\acknowledgments

GS acknowledges T.Giamarchi and P. Le Doussal for stimulating
discussions. The author's
financial support is provided through the 
European Community's Human Potential Program under contract
HPRN-CT-2002-00307, DYGLAGEMEM.


\begin{thebibliography}{23}
\expandafter\ifx\csname natexlab\endcsname\relax\def\natexlab#1{#1}\fi
\expandafter\ifx\csname bibnamefont\endcsname\relax
  \def\bibnamefont#1{#1}\fi
\expandafter\ifx\csname bibfnamefont\endcsname\relax
  \def\bibfnamefont#1{#1}\fi
\expandafter\ifx\csname citenamefont\endcsname\relax
  \def\citenamefont#1{#1}\fi
\expandafter\ifx\csname url\endcsname\relax
  \def\url#1{\texttt{#1}}\fi
\expandafter\ifx\csname urlprefix\endcsname\relax\def\urlprefix{URL }\fi
\providecommand{\bibinfo}[2]{#2}
\providecommand{\eprint}[2][]{\url{#2}}

\bibitem[{\citenamefont{Zeller and Pohl}(1971)}]{zeller_chalspe_struct_glasses}
\bibinfo{author}{\bibfnamefont{R.~C.} \bibnamefont{Zeller}} \bibnamefont{and}
  \bibinfo{author}{\bibfnamefont{R.~O.} \bibnamefont{Pohl}},
  \bibinfo{journal}{Phys. Rev. B} \textbf{\bibinfo{volume}{4}},
  \bibinfo{pages}{2029} (\bibinfo{year}{1971}).

\bibitem[{\citenamefont{Ackerman et~al.}(1981)\citenamefont{Ackerman, Moy,
  Potter, and Anderson}}]{ackerman_chalspe_disordered_crystals}
\bibinfo{author}{\bibfnamefont{D.~A.} \bibnamefont{Ackerman}},
  \bibinfo{author}{\bibfnamefont{D.}~\bibnamefont{Moy}},
  \bibinfo{author}{\bibfnamefont{R.~C.} \bibnamefont{Potter}},
  \bibnamefont{and} \bibinfo{author}{\bibfnamefont{A.~C.}
  \bibnamefont{Anderson}}, \bibinfo{journal}{Phys. Rev. B}
  \textbf{\bibinfo{volume}{23}}, \bibinfo{pages}{3886} (\bibinfo{year}{1981}),
  \bibinfo{note}{and references therein}.

\bibitem[{bin()}]{binder_spinglass_review}
\bibinfo{note}{For a review see K. Binder, A. P. Young, Rev. Mod. Phys. {\bf
  58} 801 (1986)}.

\bibitem[{\citenamefont{Anderson et~al.}(1972)\citenamefont{Anderson, Halperin,
  and Varma}}]{anderson_twolevels}
\bibinfo{author}{\bibfnamefont{P.~W.} \bibnamefont{Anderson}},
  \bibinfo{author}{\bibfnamefont{B.~I.} \bibnamefont{Halperin}},
  \bibnamefont{and} \bibinfo{author}{\bibfnamefont{C.~M.} \bibnamefont{Varma}},
  \bibinfo{journal}{Phil. Mag.} \textbf{\bibinfo{volume}{25}},
  \bibinfo{pages}{1} (\bibinfo{year}{1972}).

\bibitem[{\citenamefont{Bray and Moore}(1980)}]{bray_quantum_sg}
\bibinfo{author}{\bibfnamefont{A.~J.} \bibnamefont{Bray}} \bibnamefont{and}
  \bibinfo{author}{\bibfnamefont{M.~A.} \bibnamefont{Moore}},
  \bibinfo{journal}{J. Phys. C} \textbf{\bibinfo{volume}{13}},
  \bibinfo{pages}{L655} (\bibinfo{year}{1980}).

\bibitem[{\citenamefont{Sachdev and Ye}(1993)}]{sachdev_ye_sg}
\bibinfo{author}{\bibfnamefont{S.}~\bibnamefont{Sachdev}} \bibnamefont{and}
  \bibinfo{author}{\bibfnamefont{J.}~\bibnamefont{Ye}}, \bibinfo{journal}{Phys.
  Rev. Lett.} \textbf{\bibinfo{volume}{70}}, \bibinfo{pages}{3339}
  (\bibinfo{year}{1993}).

\bibitem[{\citenamefont{Cugliandolo et~al.}(2001)\citenamefont{Cugliandolo,
  Grempel, and {da Silva Santos}}}]{cugliandolo_quantum_p_spin}
\bibinfo{author}{\bibfnamefont{L.~F.} \bibnamefont{Cugliandolo}},
  \bibinfo{author}{\bibfnamefont{D.~R.} \bibnamefont{Grempel}},
  \bibnamefont{and} \bibinfo{author}{\bibfnamefont{C.~A.} \bibnamefont{{da
  Silva Santos}}}, \bibinfo{journal}{Phys. Rev. B}
  \textbf{\bibinfo{volume}{64}}, \bibinfo{pages}{14403} (\bibinfo{year}{2001}).

\bibitem[{geo()}]{georges_mf_quantum_spinglass}
\bibinfo{note}{A. Georges, O. Parcollet and S. Sachdev , Phys. Rev. Lett. {\bf
  85}, 840 (2000); Phys. Rev. B {\bf 63}, 134406 (2001)}.

\bibitem[{\citenamefont{Giamarchi and {Le
  Doussal}}(1996)}]{giamarchi_columnar_variat}
\bibinfo{author}{\bibfnamefont{T.}~\bibnamefont{Giamarchi}} \bibnamefont{and}
  \bibinfo{author}{\bibfnamefont{P.}~\bibnamefont{{Le Doussal}}},
  \bibinfo{journal}{Phys. Rev. B} \textbf{\bibinfo{volume}{53}},
  \bibinfo{pages}{15206} (\bibinfo{year}{1996}).

\bibitem[{\citenamefont{Camjayi and Rozenberg}(2003)}]{camjayi_sun_sg_num}
\bibinfo{author}{\bibfnamefont{A.}~\bibnamefont{Camjayi}} \bibnamefont{and}
  \bibinfo{author}{\bibfnamefont{M.~J.} \bibnamefont{Rozenberg}},
  \bibinfo{journal}{Phys. Rev. Lett.} \textbf{\bibinfo{volume}{90}},
  \bibinfo{pages}{217202} (\bibinfo{year}{2003}).

\bibitem[{\citenamefont{Schehr et~al.}(2004{\natexlab{a}})\citenamefont{Schehr,
  Giamarchi, and {Le Doussal}}}]{schehr_chalspe_quantique}
\bibinfo{author}{\bibfnamefont{G.}~\bibnamefont{Schehr}},
  \bibinfo{author}{\bibfnamefont{T.}~\bibnamefont{Giamarchi}},
  \bibnamefont{and} \bibinfo{author}{\bibfnamefont{P.}~\bibnamefont{{Le
  Doussal}}}, \bibinfo{journal}{Europhys. Lett.} \textbf{\bibinfo{volume}{66}},
  \bibinfo{pages}{538} (\bibinfo{year}{2004}{\natexlab{a}}).

\bibitem[{\citenamefont{Giamarchi and {Le
  Doussal}}(1995)}]{giamarchi_vortex_long}
\bibinfo{author}{\bibfnamefont{T.}~\bibnamefont{Giamarchi}} \bibnamefont{and}
  \bibinfo{author}{\bibfnamefont{P.}~\bibnamefont{{Le Doussal}}},
  \bibinfo{journal}{Phys. Rev. B} \textbf{\bibinfo{volume}{52}},
  \bibinfo{pages}{1242} (\bibinfo{year}{1995}).

\bibitem[{\citenamefont{Gr{\"u}ner}(1994)}]{gruner_book_cdw}
\bibinfo{author}{\bibfnamefont{G.}~\bibnamefont{Gr{\"u}ner}},
  \emph{\bibinfo{title}{Density Waves in Solids}}
  (\bibinfo{publisher}{Addison-Wesley, Reading}, \bibinfo{address}{MA},
  \bibinfo{year}{1994}).

\bibitem[{\citenamefont{Giamarchi}(2002)}]{giamarchi_wigner_review}
\bibinfo{author}{\bibfnamefont{T.}~\bibnamefont{Giamarchi}}, in
  \emph{\bibinfo{booktitle}{Strongly correlated fermions and bosons in low
  dimensional disordered systems}}, edited by
  \bibinfo{editor}{\bibfnamefont{I.~V.} \bibnamefont{{Lerner et al.}}}
  (\bibinfo{publisher}{Kluwer}, \bibinfo{address}{Dordrecht},
  \bibinfo{year}{2002}), \bibinfo{note}{cond-mat/0205099}.

\bibitem[{\citenamefont{Chitra et~al.}(2001)\citenamefont{Chitra, Giamarchi,
  and {Le Doussal}}}]{chitra_wigner_long}
\bibinfo{author}{\bibfnamefont{R.}~\bibnamefont{Chitra}},
  \bibinfo{author}{\bibfnamefont{T.}~\bibnamefont{Giamarchi}},
  \bibnamefont{and} \bibinfo{author}{\bibfnamefont{P.}~\bibnamefont{{Le
  Doussal}}}, \bibinfo{journal}{Phys. Rev. B} \textbf{\bibinfo{volume}{65}},
  \bibinfo{pages}{035312} (\bibinfo{year}{2001}).

\bibitem[{\citenamefont{Blatter et~al.}(1994)\citenamefont{Blatter, Feigel'man,
  Geshkenbein, Larkin, and Vinokur}}]{blatter_vortex_review}
\bibinfo{author}{\bibfnamefont{G.}~\bibnamefont{Blatter}},
  \bibinfo{author}{\bibfnamefont{M.~V.} \bibnamefont{Feigel'man}},
  \bibinfo{author}{\bibfnamefont{V.~B.} \bibnamefont{Geshkenbein}},
  \bibinfo{author}{\bibfnamefont{A.~I.} \bibnamefont{Larkin}},
  \bibnamefont{and} \bibinfo{author}{\bibfnamefont{V.~M.}
  \bibnamefont{Vinokur}}, \bibinfo{journal}{Rev. Mod. Phys.}
  \textbf{\bibinfo{volume}{66}}, \bibinfo{pages}{1125} (\bibinfo{year}{1994}).

\bibitem[{\citenamefont{Giamarchi and {Le
  Doussal}}(1998)}]{giamarchi_book_young}
\bibinfo{author}{\bibfnamefont{T.}~\bibnamefont{Giamarchi}} \bibnamefont{and}
  \bibinfo{author}{\bibfnamefont{P.}~\bibnamefont{{Le Doussal}}},
  \emph{\bibinfo{title}{Statics and dynamics of disordered elastic systems}}
  (\bibinfo{publisher}{World Scientific}, \bibinfo{address}{Singapore},
  \bibinfo{year}{1998}), p. \bibinfo{pages}{321},
  \bibinfo{note}{cond-mat/9705096}.

\bibitem[{\citenamefont{Nattermann and
  Scheidl}(2000)}]{nattermann_vortex_review}
\bibinfo{author}{\bibfnamefont{T.}~\bibnamefont{Nattermann}} \bibnamefont{and}
  \bibinfo{author}{\bibfnamefont{S.}~\bibnamefont{Scheidl}},
  \bibinfo{journal}{Adv. Phys.} \textbf{\bibinfo{volume}{49}},
  \bibinfo{pages}{607} (\bibinfo{year}{2000}).

\bibitem[{\citenamefont{Giamarchi and
  Bhattacharya}(2002)}]{giamarchi_vortex_review}
\bibinfo{author}{\bibfnamefont{T.}~\bibnamefont{Giamarchi}} \bibnamefont{and}
  \bibinfo{author}{\bibfnamefont{S.}~\bibnamefont{Bhattacharya}}, in
  \emph{\bibinfo{booktitle}{High Magnetic Fields: Applications in Condensed
  Matter Physics and Spectroscopy}}, edited by
  \bibinfo{editor}{\bibfnamefont{C.}~\bibnamefont{{Berthier et al.}}}
  (\bibinfo{publisher}{Springer-Verlag}, \bibinfo{address}{Berlin},
  \bibinfo{year}{2002}), p. \bibinfo{pages}{314},
  \bibinfo{note}{cond-mat/0111052}.

\bibitem[{\citenamefont{M{\'e}zard and
  Parisi}(1991)}]{mezard_variational_replica}
\bibinfo{author}{\bibfnamefont{M.}~\bibnamefont{M{\'e}zard}} \bibnamefont{and}
  \bibinfo{author}{\bibfnamefont{G.}~\bibnamefont{Parisi}},
  \bibinfo{journal}{J. de Phys. I} \textbf{\bibinfo{volume}{1}},
  \bibinfo{pages}{809} (\bibinfo{year}{1991}).

\bibitem[{\citenamefont{Schehr et~al.}(2003)\citenamefont{Schehr, Giamarchi,
  and {Le Doussal}}}]{schehr_chalspe_classique}
\bibinfo{author}{\bibfnamefont{G.}~\bibnamefont{Schehr}},
  \bibinfo{author}{\bibfnamefont{T.}~\bibnamefont{Giamarchi}},
  \bibnamefont{and} \bibinfo{author}{\bibfnamefont{P.}~\bibnamefont{{Le
  Doussal}}}, \bibinfo{journal}{Phys. Rev. Lett.}
  \textbf{\bibinfo{volume}{91}}, \bibinfo{pages}{117002}
  (\bibinfo{year}{2003}).

\bibitem[{\citenamefont{Schehr et~al.}(2004{\natexlab{b}})\citenamefont{Schehr,
  Giamarchi, and {Le Doussal}}}]{schehr_chalspe_long}
\bibinfo{author}{\bibfnamefont{G.}~\bibnamefont{Schehr}},
  \bibinfo{author}{\bibfnamefont{T.}~\bibnamefont{Giamarchi}},
  \bibnamefont{and} \bibinfo{author}{\bibfnamefont{P.}~\bibnamefont{{Le
  Doussal}}} (\bibinfo{year}{2004}{\natexlab{b}}),
  \bibinfo{note}{cond-mat/0411227}.

\bibitem[{\citenamefont{L.F.Cugliandolo
  et~al.}(2004)\citenamefont{L.F.Cugliandolo, Giamarchi, and {Le
  Doussal}}}]{cugliandolo_keldysh_elastic}
\bibinfo{author}{\bibnamefont{L.F.Cugliandolo}},
  \bibinfo{author}{\bibfnamefont{T.}~\bibnamefont{Giamarchi}},
  \bibnamefont{and} \bibinfo{author}{\bibfnamefont{P.}~\bibnamefont{{Le
  Doussal}}} (\bibinfo{year}{2004}), \bibinfo{note}{in preparation.}

\end{thebibliography}

\appendix

\section{Low temperature expansion for Model I and Model II : detailed
  calculations.}\label{app_low_T_gen_BG}

In this appendix, we show some general properties of the low
temperature expansion of multiple sums over Matsubara
frequencies. We will use here the notations of the elastic manifold
(\ref{Eq_I}-\ref{wc}) (the extension to the quantum $p$-spin model
being straighforward). 

\subsection{A first stage with multiple Matsubara sums.}

To begin with, we restrict ourselves to the case where $\Sigma$ and
$I(\omega_n)$ as a function of $\omega_n$ do not depend on temperature
$T$, for all $\omega_n$ including the mode
$\omega_n = 0$. We also assume the low frequency behavior $I(\omega_n)
\sim |\omega_n| + {\cal O}(\omega_n^2)$ (\ref{Eq_I}). 
Here we are interested
in the low temperature 
expansion of the following quantity (\ref{Property_1_bg}):
\begin{eqnarray}
\int_0^{\beta \hbar} d\tau \left ({\cal H}(\tilde{B}(\tau)) - {\cal
  H}(B)\right ) 
\end{eqnarray}  
As we will see, the first non vanishing finite temperature correction
  is {\it a 
  priori} of order $T^2$ : we show that this contribution in
  fact cancels for any (smooth enough) function ${\cal H}(X)$.  
To show this cancellation, we show the following property, for any
  integer $m$:
\begin{eqnarray}\label{gen_id_1}
\int_0^{\beta\hbar} d\tau
\left(\tilde{B}(\tau)^m - B^m\right)  = C^{\text{st}} + {\cal O}(T^4) 
\end{eqnarray}
We introduce de notation
\begin{eqnarray}
K(\omega_n) = J_1(M \omega_n^2 + \Sigma + I(\omega_n))
\end{eqnarray}
Inserting the definitions of $B$ (\ref{def_B}) and
$\tilde{B}(\tau)$ (\ref{def_b_tau}) in (\ref{gen_id_1}), one obtains
\begin{eqnarray}
&&\int_0^{\beta\hbar} d\tau \left(\tilde{B}(\tau)^m - B^m\right) = \\
&&\frac{(2\hbar)^m}{(\beta\hbar)^m} \sum_{n_1,..,n_m} \sum_{k=1}^m (-1)^k
  C^k_m \int_0^{\beta\hbar} d\tau
  \cos{(\omega_{n_1}\tau)}..\cos{(\omega_{n_k}\tau)} \nonumber \\
&& \times
  K(\omega_{n_1})..
  K(\omega_{n_k})K(\omega_{n_{k+1}}).. K(\omega_{n_m}) \nonumber    
\end{eqnarray}
where $C^k_m = m!/((m-k)!k!)$.
Performing the integral over $\tau$, using the property of parity 
$K(\omega_n) = K(-\omega_n)$, this expression can be written as
\begin{eqnarray}\label{expansion_1}
&&\int_0^{\beta\hbar} d\tau \left( \tilde{B}(\tau)^m - B^m \right)  = -2\hbar
  J_1(\Sigma) m B^{m-1} \\ 
&&+ 
\sum_{k=2}^{m}(-1)^k C^k_m \frac{(2\hbar)^k}{(\beta\hbar)^{k-1}}
  \sum_{n_1,..,n_{k-1}} K(\omega_{n_1})..K(\omega_{n_{k-1}})
  \nonumber \\
&&\times  K(\omega_{n_1}+...+\omega_{n_{k-1}}) B^{m-k}  \nonumber
\end{eqnarray} 
We now focus on the low temperature expansion of the multiple sum over
Matsubara frequencies. In that purpose, we use the spectral
representation of the Green's funtion : 
\begin{eqnarray}\label{Def_spectral}
&&\frac{1}{cq^2 + M\omega_n^2 + \Sigma + I(\omega_n)} = -
\int_{-\infty}^{\infty} \frac{d\omega}{\pi} A(q,\omega) \frac{1}{i\omega_n -
  \omega} \nonumber \\
&& A(q,\omega) =
\frac{I''(\omega)}{(cq^2- M\omega^2+\Sigma+I'(\omega))^2 + (I''(\omega))^2} 
\end{eqnarray}
where the spectral function $A(q,\omega)$ is the imaginary part of the
retarded function and 
we remind $I(\omega_n \to -i\omega + 0^+) = I'(\omega) + i I''(\omega)$. 
All the temperature dependence  
is then contained in the different Bose factors: 
\begin{eqnarray}\label{structure}
&&\hspace*{-0.5cm} \frac{1}{(\beta\hbar)^{k-1}}\sum_{n_1,..,n_{k-1}}
  K(\omega_{n_1})..K(\omega_{n_{k-1}}) 
  K(\omega_{n_1}+...+\omega_{n_{k-1}}) \nonumber \\
&&\hspace*{-0.5cm} = \frac{(-1)^k}{(\pi)^k} \int d\epsilon_1..d\epsilon_k
  {\cal A}(\epsilon_1) .. {\cal A}(\epsilon_k) \\
&&\hspace*{-0.5cm}\times
  \frac{1}{(\beta\hbar)^{k-1}}\sum_{n_1,..,n_{k-1}}\frac{1}{i\omega_{n_1}
  - 
  \epsilon_1}...\frac{1}{i\omega_{n_1} + .. + 
  i\omega_{n_{k-1}} - \epsilon_k} \nonumber \\
&&\hspace*{-0.5cm} = \frac{1}{\pi^k}\int d\epsilon_1..d\epsilon_k
  {\cal A}(\epsilon_1) .. {\cal A}(\epsilon_k)(f_B(\epsilon_1) -
  f_B(\epsilon_k))\\
&&\hspace*{-0.5cm} \times (f_B(\epsilon_2) -
  f_B(\epsilon_k-\epsilon_1))(f_B(\epsilon_3) - 
  f_B(\epsilon_k-\epsilon_1-\epsilon_2)) .. \nonumber \\
&&\hspace*{-0.5cm} \times \frac{(f_B(\epsilon_{k-1}) -
  f_B(\epsilon_k - \epsilon_1-\epsilon_2-..-\epsilon_{k-2}))}
  {\epsilon_k - (\epsilon_1 + ..+\epsilon_{k-1})}      
\end{eqnarray}
where ${\cal A}(\omega) = \int_q A(q,\omega)$ and $f_B(\epsilon)$
the Bose factor. This expression 
(\ref{structure}) has a very interesting structure which allows us to
extract simply the term of order $T^2$. Indeed, considering the
low temperature expansion of the following term (which is the
analogous of a Sommerfeld expansion in the fermionic case) for any
function ${\cal H}(x)$ whith ${\cal H}(\mu) = 0$ 
\begin{eqnarray}\label{sommerfeld}
&&\int d\epsilon H(\epsilon) f_B(\epsilon-\mu) = -\int_{-\infty}^{\mu}
d\epsilon H(\epsilon) \nonumber\\
&& + \left(\frac{T}{\hbar}\right)^2  \frac{\pi^2}{3}H'(\mu) +
{\cal O}(T^4)  
\end{eqnarray}    
notice of course that the asumption ${\cal H}(\mu =0)$ is of course crucial
here (\ref{sommerfeld}).  
This expansion (\ref{sommerfeld}) allows us to write
formally the Bose factor, when inserted in an integral over frequency
$\epsilon$: 
\begin{equation}
f_B(\epsilon-\mu) \equiv -\theta(-\epsilon + \mu) +
\left(\frac{T}{\hbar}\right)^2 
\frac{\pi^2}{3} \delta(\epsilon-\mu) \partial_{\epsilon} +
{\cal O}(T^4) \label{lowT_Bose}  
\end{equation}
where $\theta(x)$ is the step function ($\theta(x) = 1$ for $x>0$,
$\theta(x)=0$ if $x<0$), and 
the notation $\partial_{\epsilon}$ stands for a derivative acting on
the function which enters multiplicatively with $f_B(\epsilon)$ the
integral over $\epsilon$. This form (\ref{lowT_Bose}) is very suitable
to extract the coefficient of the term  of order ${\cal O}(T^2)$ in
(\ref{structure}). Indeed, using (\ref{lowT_Bose}), one discovers 
in (\ref{structure}), that only the terms where the Bose factors have
{\it only one frequency} in their argument do contribute to order
${\cal O}(T^2)$. If we expand, for instance
$f_B(\epsilon_k-\epsilon_1-\epsilon_2)$ in (\ref{structure}) one
obtains, for $k \geq 3$
\begin{eqnarray}
&&\hspace*{-0.5cm}\frac{1}{(\beta\hbar)^2} \frac{\pi^2}{3}\frac{1}{\pi^k}\int
  d\epsilon_1..d\epsilon_k 
  \delta(-\epsilon_k + \epsilon_1 + \epsilon_2)\partial_{\epsilon_k}
  \Big [ {\cal A}(\epsilon_1) 
  .. {\cal A}(\epsilon_k) \nonumber \\
&&\hspace*{-0.5cm}\times (-\theta(-\epsilon_1) + 
  \theta(-\epsilon_k)) \nonumber (-\theta(-\epsilon_2) +
  \theta(-\epsilon_k+\epsilon_1)).. \nonumber \\
&&\hspace*{-0.5cm}\times
\frac{(-\theta(-\epsilon_{k-1}) +
  \theta(-\epsilon_k + \epsilon_1+\epsilon_2+..+\epsilon_{k-2}))}
  {\epsilon_k - (\epsilon_1 + ..+\epsilon_{k-1})} \Big ] 
\end{eqnarray}
Because of the function $\delta(\epsilon_k - \epsilon_1 -
\epsilon_2)$, one must take the derivative of the terms
 $\partial_{\epsilon_k}
(-\theta(-\epsilon_2) + \theta(-\epsilon_k+\epsilon_1) =
-\delta(-\epsilon_k + \epsilon_1)$ (which is a priori the only non
vanishing term), which yields 
\begin{eqnarray}
&&\frac{1}{(\beta\hbar)^2} \frac{\pi^2}{3}\frac{1}{\pi^k}\int
  d\epsilon_1..d\epsilon_k 
  \delta(-\epsilon_k + \epsilon_1 + \epsilon_2)\delta(-\epsilon_k +
  \epsilon_1) \nonumber \\
&&\times {\cal A}(\epsilon_1) .. {\cal A}(\epsilon_k)(-\theta(-\epsilon_1) + 
  \theta(-\epsilon_k)).. \\
&&\times \frac{(-\theta(-\epsilon_{k-1}) +
  \theta(-\epsilon_k + \epsilon_1+\epsilon_2+..+\epsilon_{k-2}))}
  {\epsilon_k - (\epsilon_1 + ..+\epsilon_{k-1})}  = 0 \nonumber
\end{eqnarray}
We can easily generalise this mechanism to Bose factors having more
than $3$ frequencies in their arguments. The case of Bose factors
having $2$ frequencies in their argument has to be treated
separately. This one, i.e. $f_B(\epsilon_k - \epsilon_1)$ in
(\ref{structure})) yields
\begin{eqnarray}
&&\frac{1}{(\beta\hbar)^2} \frac{\pi^2}{3}\frac{1}{\pi^k}\int
  d\epsilon_1..d\epsilon_k \delta(-\epsilon_k + \epsilon_1)
  \partial_{\epsilon_k} \Big[ {\cal A}(\epsilon_1) 
  .. {\cal A}(\epsilon_k) \nonumber \\
&&\times (-\theta(-\epsilon_1) + 
  \theta(-\epsilon_k))(-\theta(-\epsilon_3) + \theta(-\epsilon_k +
  \epsilon_1 + \epsilon_2).. \nonumber \\
&&\times \frac{(-\theta(-\epsilon_{k-1}) +
  \theta(-\epsilon_k + \epsilon_1+\epsilon_2+..+\epsilon_{k-2}))}
  {\epsilon_k - (\epsilon_1 + ..+\epsilon_{k-1})} \Big] 
\end{eqnarray} 
Here again, one has to take the derivative 
$\partial_{\epsilon_k}(-\theta(-\epsilon_1) + \theta(-\epsilon_k)) =
-\delta(-\epsilon_k)$, which yields
\begin{eqnarray}
&&\frac{1}{(\beta\hbar)^2} \frac{\pi^2}{3}\frac{1}{\pi^k}\int
  d\epsilon_1..d\epsilon_k \delta(-\epsilon_k +
  \epsilon_1)\delta(-\epsilon_k) 
  {\cal A}(\epsilon_1) 
  .. {\cal A}(\epsilon_k) \nonumber \\
&&\times(-\theta(-\epsilon_3) + \theta(-\epsilon_k +
  \epsilon_1 + \epsilon_2))..\nonumber \\
&&\times \frac{(-\theta(-\epsilon_{k-1}) +
  \theta(-\epsilon_k + \epsilon_1+\epsilon_2+..+\epsilon_{k-2}))}
 {\epsilon_k - (\epsilon_1 + ..+\epsilon_{k-1})} \Big] \nonumber \\
&& = 0
\end{eqnarray} 
where we have used ${\cal A}(0) = 0$. We conclude from this analysis that only the
Bose factors having only one frequency in their argument, such as
$f_B(\epsilon_1), f_B(\epsilon_2)..$ do contribute to the coefficient
of order $T^2$ in (\ref{structure}). Notice that they all give
the same contribution, thus the low temperature expansion of   
(\ref{structure}) can be written for $k\geq 3$
\begin{eqnarray}
&&\hspace*{-0.5cm}\frac{1}{(\beta\hbar)^{k-1}}\sum_{n_1,..,n_{k-1}}
  K(\omega_{n_1})..K(\omega_{n_{k-1}}) 
  K(\omega_{n_1}+..+\omega_{n_{k-1}}) \nonumber \\
&&\hspace*{-0.5cm} = {\cal J}_k + \frac{k}{(\beta\hbar)^2} \frac{\pi}{3} {\cal A}'(0) {\cal
  J}_{k-1} + 
  O(\beta^{-4}) 
  \nonumber \\
&&\hspace*{-0.5cm}{\cal J}_k = 
\frac{1}{\pi^k}\int d\epsilon_1..d\epsilon_k
  {\cal A}(\epsilon_1) .. {\cal A}(\epsilon_k)(-\theta(-\epsilon_1) +
  \theta(-\epsilon_k)) \nonumber \\
&&\hspace*{-0.5cm}\times (-\theta(-\epsilon_2) + \theta
  (-\epsilon_k+\epsilon_1))(-\theta(-\epsilon_3) +
  \theta(-\epsilon_k+\epsilon_1+\epsilon_2)) \nonumber \\
&&\hspace*{-0.5cm}\times..\frac{(-\theta(-\epsilon_{k-1}) +
  \theta(-\epsilon_k + \epsilon_1+\epsilon_2+..+\epsilon_{k-2}))}
  {\epsilon_k - (\epsilon_1 + ..+\epsilon_{k-1})}  
\end{eqnarray}
where we have used ${\cal A}(-\epsilon) = -{\cal A}(\epsilon)$ to treat the term 
$f_B(\epsilon_k)$ in (\ref{structure}). For the particular case $k=2$, one has 
\begin{eqnarray}
&&\hspace*{-0.3cm}\frac{1}{\beta\hbar} \sum_{n_1} K(\omega_{n_1})^2 = {\cal J}_2 +
\frac{2}{(\beta\hbar)^2}\frac{{\cal A}'(0)}{3} \int d\epsilon
\frac{{\cal A}(\epsilon)}{\epsilon} + {\cal O}(T^4)\nonumber \\
&& \hspace*{-0.3cm} = {\cal J}_2 +
\frac{2}{(\beta\hbar)^2} K(\Sigma) \frac{\pi}{3} {\cal A}'(0) + O(\beta^{-4})
\end{eqnarray}
where we have used $\int d\epsilon {\cal A}(\epsilon)/\epsilon = \pi
J_1(\Sigma)$. Finally, we need the low $T$ expansion of $B$ (\ref{def_B}):
\begin{eqnarray}\label{devel_B}
B = B^{(0)} + {2\hbar} \left(\frac{T}{\hbar}\right)^2 \frac{\pi}{3}
{\cal A}'(0) + {\cal O}(T^4)
\end{eqnarray}
Using these behaviors, one obtains:
\begin{widetext}
\begin{eqnarray}\label{last_step}
&&\int_0^{\beta\hbar} d\tau \left(
\tilde{B}(\tau)^m - B^m\right)   = C^{\text{st}} +
\left(\frac{T}{\hbar}\right)^2 \Big(  - 
  (2\hbar)^2 J_1(\Sigma) {\cal A}'(0) \frac{\pi}{3} m(m-1) 
  (B^{(0)})^{m-2} \\ 
&& + (2\hbar)^2 2 C^2_m J_1(\Sigma) {\cal A}'(0) \frac{\pi}{3} (B^{(0)})^{m-2}
 +
(2\hbar)^3 C^2_m {\cal J}_2 (m-2) {\cal A}'(0) \frac{\pi}{3} (B^{(0)})^{m-3}
\nonumber \\
&&+ \sum_{k=3}^m (2\hbar)^k (-1)^k C^k_m k
\frac{\pi}{3}{\cal A}'(0){\cal J}_{k-1} (B^{(0)})^{m-k} 
+ \sum_{k=3}^{m-1}(2\hbar)^{k+1} (-1)^k C^k_m (m-k)\frac{\pi}{3}{\cal A}'(0)
     {\cal J}_{k} 
(B^{(0)})^{m-k-1} \Big)  + {\cal O}(T^4) \nonumber
\end{eqnarray}
\end{widetext}
First, we notice that the two first terms $\propto T^2$ just cancel and
moreover 
\begin{eqnarray}\label{magic_rel}
&&\sum_{k=3}^m (2\hbar)^k (-1)^k C^k_m ( k
\frac{\pi}{3}{\cal A}'(0){\cal J}_{k-1} (B^{(0)})^{m-k} \nonumber \\
&&+ \sum_{k=3}^{m-1}(2\hbar)^k (-1)^k C^k_m (m-k)(2\hbar)\frac{\pi}{3}{\cal A}'(0)
     {\cal J}_{k} 
(B^{(0)})^{m-k-1}) \nonumber \\
&& =  \frac{\pi}{3} {\cal A}'(0)\Big(
-(2\hbar)^3 3 C^3_m {\cal J}_{2}
(B^{(0)})^{m-3} 
\nonumber \\
&& + \sum_{k=4}^m (-1)^k (2\hbar)^k {\cal J}_{k-1}
(B^{(0)})^{m-k}(k C^k_m \nonumber \\
&& - (m-k+1)C^{k-1}_m) \Big)\nonumber \\
&& = - (2\hbar)^3 \frac{m(m-1)(m-2)}{2} \frac{\pi}{3} {\cal A}'(0){\cal J}_{2}
(B^{(0)})^{m-3}    
\end{eqnarray}
This identity (\ref{magic_rel}) combined with (\ref{last_step})
yields finally to the announced property
\begin{eqnarray}\label{app_property_1}
\int_0^{\beta\hbar} d\tau
\left(\tilde{B}(\tau)^m - B^m\right)  = C^{\text{st}} + {\cal O}(T^4)
\end{eqnarray}
as announced in the text (\ref{Property_1_bg}).

\subsection{Handling the peculiar term $\propto (1-\delta_{n,0})$.}

We now consider the extension of this property (\ref{app_property_1})
to the case where $\Sigma + I(\omega_n)$ now depend on $T$, and are of
the form (\ref{devel_variat_BG}). We want to follow the same steps as
previously (\ref{app_property_1}), and use the spectral representation
of the Green 
function. Therefore, we need in principle to know the analytical
continuation of the term $\propto(1-\delta_{n,0})$ in
(\ref{devel_variat_BG}). In order to avoid this ambiguity, 
we start by isolating the term $\propto (1-\delta_{n,0})$ in
(\ref{devel_variat_BG}): 
\begin{eqnarray}\label{eqvar_T_finie_app}
&&\Sigma + I(\omega_n) = \Sigma + C(1 - \delta_{n,0}) +
\tilde{I}(\omega_n) \\
&&\tilde{I}(\omega_n) \sim |\omega_n| + {\cal O}(\omega_n^2) \nonumber \\
&&\Sigma = \Sigma^{(0)} + T^2 \Sigma^{(2)} + T^3 \Sigma^{(3)}  +
{\cal O}(T^4) \nonumber \\ 
&&C =  - T^2 \Sigma^{(2)} - T^3 \Sigma^{(3)}  +
{\cal O}(T^4) \nonumber \\
&&\tilde{I}(\omega_n) = \tilde{I}^{0}(\omega_n) + {\cal O}(T^4) \nonumber  
\end{eqnarray} 
where $\tilde{I}(\omega_n)$ is defined such that there is no more
ambiguity concerning its analytical continuation.  
%
%
We want to study the $T$ dependence of integrals over imaginary time
such as (\ref{expansion_1}) when the solution of the variational
equations are of the form (\ref{devel_variat_BG}). In order to extract
the quadratic and 
cubic terms in this expansion, we follow the previous analysis, except
that, here, the mode $\omega_n=0$ has to be
treated separately (\ref{devel_variat_BG}). Notice however that this
ambiguity does not exist for the computation of $\tilde{B}(\tau)$
(\ref{def_b_tau}).    
We analyse the low $T$ behavior of (\ref{expansion_1})
in the following way:
\begin{eqnarray}
&&\int_0^{\beta \hbar} d\tau \left( \tilde{B}(\tau)^m -
  B^m \right) = \frac{1}{\hbar}\int_0^{\beta \hbar} d\tau \left(
  \tilde{B}(\tau)^m - 
  \hat{B}^m \right) \nonumber \\
&& + \beta \hbar(\hat{B}^m - B^m) \label{isolate_zero_BG} \\  
&&\hat B = \frac{2}{\beta} \sum_n J_1(\Sigma + C + M\omega_n^2 +
  \tilde{I}(\omega_n)) \label{def_hat_B}
\end{eqnarray} 
Using the previous property (\ref{app_property_1}), one has simply
\begin{eqnarray}
\frac{1}{\hbar}\int_0^{\beta \hbar} d\tau \left( \tilde{B}(\tau)^m -
  \hat{B}^m \right) = C^{\text{st}} + {\cal O}(T^4)
\end{eqnarray}
and the leading corrections at finite temperature are then given by
the second term in (\ref{isolate_zero_BG}):
\begin{eqnarray}
&&\beta(\hat{B}^m - B^m) \\
&&= - \beta \hbar
\sum_{k=1}^m C^m_k \left(\frac{2}{\beta} \int_q
\frac{C}{(cq^2 + \Sigma)(cq^2 + \Sigma + C)}      \right)^k
\hat{B}^{m-k} \nonumber \\
&& = 2 \hbar \left(   T^2 \Sigma^{(2)} + T^3\Sigma^{(3)}\right)
J_2(\Sigma^{(0)}) m (B^{(0)})^{m-1} + {\cal O}(T^4) \nonumber
\end{eqnarray}
since $B^{(0)} = {\hat B}^{(0)}$, 
and more generally, one can write, for any $\omega_n$ and any function
${\cal H}(X)$ the generalization of 
(\ref{app_property_1}) to this 
case (\ref{devel_variat_BG})
\begin{eqnarray}\label{generalisation}
&&\int_0^{\beta \hbar} d\tau \cos{(\omega_n \tau)}\left(
  {\cal H}(\tilde{B}(\tau)) - 
  {\cal H}(B) \right)  = C^{\text{st}} \\
&& + 2\hbar \delta_{n,0} \left( T^2 \Sigma^{(2)} + T^3 \Sigma^{(3)}\right)
J_2(\Sigma^{(0)}){\cal H}'(B^{(0)}) + {\cal O}(T^4)\nonumber
\end{eqnarray}
as quoted in the text (\ref{Property_BG_2}).

%
%

\section{Heisenberg spin glass : low temperature
  expansion.}\label{app_low_T_gen_suN} 

\subsection{General properties.}

We present here the detailed analysis of the low temperature
behavior of integrals over imaginary time of the form (\ref{H_sun}): 
\begin{eqnarray}\label{int_1_sun}
{\cal I} = \int_0^{\beta} d\tau [(\tilde{\cal G}(\tau) -
g)^2(\tilde{\cal G}(-\tau) - g)^2 - g^4]
\end{eqnarray} 
We first consider the case where the Green's function
$\tilde{\cal G}(i\nu_n)$, as a function of $i \nu_n$ is independent of
$T$. Similarly to the analysis performed in Appendix
\ref{app_low_T_gen_BG}, we develop the integrand in
(\ref{int_1_sun}) and perform the integral over $\tau$ : we are then
left we multiple sums over Matsubara frequencies. Depending on the
number of Green's functions they involve, the integral over $\tau$
generates $4$ different types of terms:  
\begin{eqnarray}
&&{\cal I} = \sum_{i=1}^4 {\cal I}_i  \\
&&{\cal I}_1 = \frac{1}{\beta^3} \sum_{i\nu_1,i\nu_2,i\nu_3}
\tilde{\cal G}(i\nu_1)\tilde{\cal G}(i\nu_2)\tilde{\cal G}(i\nu_3) \tilde{\cal G}(i\nu_1
+i\nu_2 - i\nu_3) \nonumber \\
&&{\cal I}_2 = -2 g \frac{1}{\beta^2} \sum_{i\nu_1,i\nu_2}
[\tilde{\cal G}(i\nu_1)\tilde{\cal G}(i\nu_2)\tilde{\cal G}(i\nu_1+i\nu_2) \nonumber \\
&& + \tilde{\cal G}(i\nu_1)\tilde{\cal G}(i\nu_2)\tilde{\cal G}(i\nu_1-i\nu_2)   
     ] \nonumber \\
&&{\cal I}_3 = 2g^2 \frac{1}{\beta}\sum_{i\nu_1}[
2\tilde{\cal G}(i\nu_1)^2 +   \tilde{\cal G}(i\nu_1)\tilde{\cal G}(-i\nu_1)]
\nonumber \\
&&{\cal I}_4 = - 4g^3 \tilde{\cal G}(i\nu_n = 0) \nonumber 
\end{eqnarray}
We follow the standard analysis and use the spectral representation of
the Green's function to handle these terms:
\begin{eqnarray}
&&\tilde{\cal G}(i\nu_n) = - \int_{-\infty}^{\infty}\frac{d\omega}{\pi}
\rho(\omega) \frac{1}{i\nu_n - \omega} \\
&&\rho(\omega) = - \text{Im}\, \tilde{\cal G}(i\nu_n \to \omega + i0^+) 
\end{eqnarray}

Performig then the sums over
the Matsubara frequencies, we are left with the same kind of structure
as found in the previous models (\ref{structure}): 
\begin{eqnarray}
&&{\cal I}_1 = \frac{1}{\pi^4} \int d\epsilon_1
d\epsilon_2 d\epsilon_3 d\epsilon_4 \prod_{i=1}^4 \rho(\epsilon_i)
(f_B(\epsilon_1)-
f_B(\epsilon_4)) \nonumber \\
&&(f_B(\epsilon_2)-
f_B(\epsilon_4 - \epsilon_1)  ) \frac{(f_B(\epsilon_3)-
f_B(\epsilon_1 + \epsilon_2 - \epsilon_4))}{\epsilon_4
+\epsilon_3 + \epsilon_2 - \epsilon_1}  \nonumber \\
\end{eqnarray} 
Under this form, we analyse straighforwardly the low temperature
behavior, using the property demonstrated previously that only Bose
factors with {\it one} frequency in their argument do contribute to this
sum. Using the expansion (\ref{sommerfeld}), we obtain up to order
${\cal O}(T^4)$:
\begin{eqnarray}\label{devel_i1}
&&{\cal I}_1 = \frac{1}{\pi^4} \int d\epsilon_1
d\epsilon_2 d\epsilon_3 d\epsilon_4 \prod_{i=1}^4 \rho(\epsilon_i)
(-\theta(-\epsilon_1)+
\theta(-\epsilon_4))\nonumber  \\
&&(-\theta(-\epsilon_2)+
\theta(-\epsilon_4 + \epsilon_1)) \frac{(-\theta(-\epsilon_3)+
\theta(-\epsilon_1 - \epsilon_2 + \epsilon_4))}{\epsilon_4
+\epsilon_3 +\epsilon_2 - \epsilon_1} \nonumber \\
&&+T^2\frac{4}{3\pi^2}\rho'(0)\int d\epsilon_1
d\epsilon_2 d\epsilon_3
\rho(\epsilon_1)\rho(\epsilon_2)\rho(\epsilon_3)\nonumber \\
&&\times (-\theta(-\epsilon_1) +
\theta(-\epsilon_3))\frac{(-\theta(-\epsilon_2) 
+ \theta(-\epsilon_1 + \epsilon_3))}{\epsilon_3 + \epsilon_2 -
\epsilon_1}  
\end{eqnarray}
where we have used the relation (which is valid although
$\rho(\epsilon)$ is not an odd function):
\begin{eqnarray}\label{sym_rel}
&&\hspace*{-0.3cm}\int d\epsilon_1 d\epsilon_2 d\epsilon_3
\rho(\epsilon_1)\rho(\epsilon_2)( \rho(\epsilon_3) + \rho(-\epsilon_3)
)   (\theta(-\epsilon_1) - 
\theta(-\epsilon_3))\nonumber \\
&&\hspace*{-0.3cm} \times \frac{(-\theta(-\epsilon_2) 
+ \theta(-\epsilon_1 + \epsilon_3))}{\epsilon_3 + \epsilon_2 -
\epsilon_1} = 0
\end{eqnarray}
Indeed, one shows this relation (\ref{sym_rel}) by noticing that the
step functions reduce the interval of integration to  
$\epsilon_3 >0,\epsilon_2 >0, \epsilon_1 < 0 $ and $\epsilon_3
<0,\epsilon_2 <0, \epsilon_1 > 0 $. A simple permutation of the
integration variables then lead to the relation (\ref{sym_rel}).

The analysis of ${\cal I}_2$ requires the low $T$ expansion of $g$:
\begin{eqnarray}\label{std_devel_g_app}
&&g = S + \int \frac{d\epsilon}{\pi} \rho(\epsilon)f_B(\epsilon) \\
&& = g^{(0)} + T^2 \frac{\pi}{3}\rho'(0) + {\cal O}(T^4) \nonumber
\end{eqnarray} 
${\cal I}_{2}$ can be written as
\begin{eqnarray}
&&{\cal I}_2 = 2g \frac{1}{\pi^3}\int d\epsilon_1 d\epsilon_2
d\epsilon_3 \rho(\epsilon_1) \rho(\epsilon_2)\rho(\epsilon_3)
\nonumber \\
&&\times[
  \frac{(f_B(\epsilon_1) - f_B(\epsilon_3)(f_B(\epsilon_2) -
    f_B(\epsilon_3 - \epsilon_1)  ) )}{\epsilon_1 + \epsilon_2 -
    \epsilon_3}  \nonumber \\
&& + \frac{(f_B(\epsilon_1) - f_B(\epsilon_3)(f_B(\epsilon_2) -
    f_B(\epsilon_1 - \epsilon_3)  ) )}{\epsilon_1 - \epsilon_2 -
    \epsilon_3}  ] 
\end{eqnarray}
from which we obtain the low $T$ expansion up to order ${\cal O}(T^4)$:
\begin{eqnarray}\label{devel_i2}
&&{\cal I}_2 = C^{\text{st}} - T^2 \frac{4}{3 \pi^2}\rho'(0) \int
  d\epsilon_1 
d\epsilon_2 d\epsilon_3
  \rho(\epsilon_1)\rho(\epsilon_2)\rho(\epsilon_3) \nonumber \\
&&\times(-\theta(-\epsilon_1) +
  \theta(-\epsilon_3))\frac{(-\theta(-\epsilon_2) 
+ \theta(-\epsilon_1 + \epsilon_3))}{\epsilon_3 + \epsilon_2 -
\epsilon_1}  \nonumber \\
&& + \frac{2g^{(0)}}{3 \pi} \frac{\rho'(0)}{\beta^2} \int d\epsilon_1
d\epsilon_2 \rho(\epsilon_1)
\rho(\epsilon_2)[4\frac{\theta(-\epsilon_1) -
    \theta(-\epsilon_2)}{\epsilon_2 - \epsilon_1} \nonumber \\
&& -
  2\frac{\theta(\epsilon_1) - \theta(-\epsilon_2)}{\epsilon_1 + \epsilon_2}] 
\end{eqnarray}  
One obtains in a similar way the expansion of ${\cal I}_3$, writing
\begin{eqnarray}
&&{\cal I}_3 = \frac{g^2}{\pi^2} \int d\epsilon_1 d\epsilon_2
\rho(\epsilon_1) \rho(\epsilon_2)[4 \frac{f_B(\epsilon_2) -
    f_B(\epsilon_1)}{\epsilon_1 - \epsilon_2} \nonumber \\
&& + 2
  \frac{f_B(\epsilon_1) - f_B(-\epsilon_2)}{\epsilon_1 + \epsilon_2}  ] 
\end{eqnarray}
form which we obtain the low $T$ expansion:
\begin{eqnarray}\label{devel_i3}
&&{\cal I}_{3} = C^{\text{st}} + T^2 \frac{2g^{(0)}}{3 \pi}
  \rho'(0)\int d\epsilon_1 d\epsilon_2
  \rho(\epsilon_1) \rho(\epsilon_2) \\
&& [-4 \frac{\theta(-\epsilon_1) -
  \theta(-\epsilon_2)}{\epsilon_2 - \epsilon_1} \nonumber + 2
  \frac{\theta(\epsilon_1) - \theta(-\epsilon_2)}{\epsilon_1 +
  \epsilon_2}] \nonumber \\
&& + T^2 4 \pi [g^{(0)}]^2 \rho'(0) \tilde{\cal G}(i\nu_n = 0)
\end{eqnarray}
where we have used $\int d\epsilon \rho(\epsilon)/\epsilon = \pi
\tilde{\cal G}(i\nu_n = 0)$.
As $\tilde{\cal G}(i \nu_n)$ does not depend on $T$, one obtains
the expansion of ${\cal I}_4$ up to order:
\begin{eqnarray}\label{devel_i4}
{\cal I}_4 = C^{\text{st}} - T^2 4 \pi [g^{(0)}]^2 \rho'(0)
\tilde{\cal G}(i \nu_n = 0) + {\cal O}(T^4)
\end{eqnarray}

Finally, collecting the quadratic contributions in (\ref{devel_i1},
\ref{devel_i2}, \ref{devel_i3}, \ref{devel_i4}) yields
\begin{equation}
{\cal I} = \int_0^{\beta} d\tau [(\tilde{\cal G}(\tau) -
g)^2(\tilde{\cal G}(-\tau) - g)^2 - g^4] = C^{\text{st}} + {\cal O}(T^4)
\end{equation}
as announced in the text (\ref{Property_1_suN}). 


Using the same method exposed previously, we study the $T$-
dependence of the integral over $\tau$ which enters the variational
equation of this model (\ref{eqvar_sun}): 
\begin{eqnarray}\label{eq_fraki}
\mathfrak{I} = \int_0^{\beta}d\tau e^{-i\nu_n \tau} [(\tilde{\cal G}(\tau)
-g)^2(\tilde{\cal G}(-\tau) -g) + g^3] 
\end{eqnarray}
As previously, after we have performed the integrals over $\tau$ in 
(\ref{eq_fraki}), we have to handle exactly the same integrals as in 
${\cal I}_2, {\cal I}_3, {\cal I}_4$. The mechanism of cancellation of
the quadratic term is again at work here (notice however that, given
that the integrand is here a polynom of degree $3$, the terms like
(\ref{sym_rel}) are not present here). This yields
\begin{eqnarray}\label{H_sun_app}
\mathfrak{I} =  C^{\text{st}} + {\cal O}(T^4)
\end{eqnarray}
as given in the text (\ref{Property_1_suN}).

\subsection{Handling the peculiar term $\propto (1-\delta_{n,0})$.}

We generalize these properties to the case where the solution of the
variational equations is of the form shown in Eq. (\ref{Devel_variat_sun}).
We use here the same strategy as presented in the Appendix
\ref{app_low_T_gen_BG}. We isolate the mode $i \nu_n =0$ as in
(\ref{isolate_zero_BG}, \ref{def_hat_B}), and given that $(\tilde{\cal
  G}(\tau) - g)$ or $(\tilde{\cal G}(-\tau) - g)$ do not depend on this
$i\nu_n=0$ mode, one obtains straightforwardly: 
\begin{eqnarray}\label{generalisation_sun}
&&{\cal I} = \int_0^{\beta \hbar} d\tau [(\tilde{\cal G}(\tau) -
g)^2(\tilde{\cal G}(-\tau) - g)^2 - g^4] = C^{\text{st}} \nonumber \\
&&-4\frac{\Theta}{J}g^{(0)} \left(T^2 g^{(2)}  +
g^{(3)} T^3 \right) + {\cal O}(T^4)  \\
&&\mathfrak{I}=\int_0^{\beta}d\tau e^{-i\nu_n \tau} [(\tilde{\cal G}(\tau)
-g)^2(\tilde{\cal G}(-\tau) -g) + g^3] = C^{\text{st}} \nonumber \\
&& + \delta_{n,0}\frac{3
\Theta}{J} \left( T^2 g^{(2)} + g^{(3)} T^3  \right) + {\cal O}(T^4)
\end{eqnarray} 
as anounced in the text (\ref{Devel_inter_sun},\ref{H_sun_devel1}).

\end{document}